\documentstyle[preprint,aps,floats,psfig,subeqn,rotate,eqsecnum]{revtex}
\def \etal {{\it et al.}} 
\def \ie {{\it i.e.}}
 
\newcommand {\bare}	{\bar{e}}

\newcommand {\bargz}	{\bar{g}_Z}
 
\newcommand {\baralpha}	{\bar{\alpha}} 
\newcommand {\baresq}	{\bar{e}^2}

\newcommand {\barssq}	{\bar{s}^2}
\newcommand {\barcsq}	{\bar{c}^2}
\newcommand {\bargzsq}	{\bar{g}_Z^2}
\newcommand {\bargwsq}	{\bar{g}_W^2} 

\newcommand {\hate}	{\hat{e}}
\newcommand {\hatg}	{\hat{g}}
\newcommand {\hats}	{\hat{s}}
\newcommand {\hatc}	{\hat{c}}
\newcommand {\hatgz}	{\hat{g}_Z}

\newcommand {\hatesq}	{\hat{e}^2}
\newcommand {\hatgsq}	{\hat{g}^2}
\newcommand {\hatssq}	{\hat{s}^2}
\newcommand {\hatcsq}	{\hat{c}^2}
\newcommand {\hatgzsq}	{\hat{g}_Z^2}

\newcommand {\mwsq}  {m_W^2}
\newcommand {\mzsq}  {m_Z^2}
\newcommand {\Lamsq} {\Lambda^2}

\newcommand {\fdw}     {f_{DW}} 
\newcommand {\fdb}     {f_{DB}}
\newcommand {\fbw}     {f_{BW}}
\newcommand {\fpone}   {f_{\Phi,1}}
\newcommand {\fwww}    {f_{WWW}}
\newcommand {\fw}      {f_{W}}
\newcommand {\fb}      {f_{B}}

\newcommand {\obw}     {{\cal O}_{BW}}
\newcommand {\opone}   {{\cal O}_{\Phi,1}}

\newcommand {\ow}      {{\cal O}_{W}}
\newcommand {\ob}      {{\cal O}_{B}}

\newcommand {\alp}[1] {\alpha_{#1}}

\newcommand {\nl}[1]     {{\cal L}_{#1}}

\newcommand {\eeww} {$e^+e^- \to W^+W^-$\/}
\newcommand {\eeff} {$e^+e^- \to f \overline{f}$}
\newcommand{\tr}{{\rm Tr}\,}

\newcommand{\cm}{\raisebox{-0.2cm}{\bf X }}
\newcommand{\cmo}{\raisebox{-0.2cm}{\bf O }}

\newcommand{\dcov}{{D}}
\newcommand{\hc}{{\rm h.c.}}
\newcommand{\id}{{\bf 1}}

\draft

\preprint{\vbox{\baselineskip14pt
\hbox{\bf KEK-TH-497}
\hbox{\bf KEK Preprint 96-135}
\hbox{\bf MPI-PhT/96-70}
\hbox{\bf NIIG-DP-96-1}
\hbox{\bf HUE-25}                 
\hbox{November 1996} }} 
\title{Probing nonstandard bosonic interactions via $W$-boson pair production 
at lepton colliders} 
\author{K.~Hagiwara$^{1,2,3}$, T.~Hatsukano$^4$, S.~Ishihara$^5$ 
and R.~Szalapski$^1$}
\address{$^1$Theory Group, KEK, Tsukuba, Ibaraki 305, Japan \\[-.2cm]
	 $^2$ICEPP, University of Tokyo, Hongo, Bunkyo-ku, Tokyo 113, 
	  Japan\\[-.1cm]
	 $^3$Max Planck Inst. f\"ur Physik, F\"ohringer Ring 6, 
	 D-80805 M\"unchen, Germany\\[-.1cm]
	 $^4$Department of Physics, Niigata University, Ikarashi, 
          Niigata 950-21, Japan\\ [-.1cm]
	 $^5$Department of Physics, Hyogo University of Education, 
         Yashiro, Hyogo 673-14, Japan}
 
\begin{document}
\maketitle 

\begin{abstract}

The process \eeww\/ provides a valuable laboratory to test the Standard Model 
(SM) and to search for new physics.  The most general helicity amplitudes for
this process require the introduction of nine form-factors which we calculate 
in the context of SU(2)$\times$U(1) gauge-invariant extensions of the SM.  The 
contributions of new physics are parametrized via an effective Lagrangian 
constructed from the light fields.  Because the mechanism of electroweak 
symmetry-breaking remains an open problem we consider both the effective 
Lagrangian with a linearly realized Higgs sector, \ie\/ with a light physical 
Higgs boson, and the effective Lagrangian which utilizes a nonlinear realization 
of the Higgs mechanism.  The use of an effective Lagrangian allows one to 
calculate consistently nonstandard contributions to \eeww\/ amplitudes as well 
as the nonstandard contributions to other processes. We study the interplay of 
the low-energy and $Z$-pole measurements with measurements via the processes 
\eeff\/ and \eeww\/ at LEP~II or a future linear $e^+e^-$  collider.  Concrete 
relationships between operators of the linear and nonlinear realizations are 
presented where possible.
\end{abstract}
\newpage
 

\section{Introduction}\label{sec-intro}

The Standard Model (SM) of electroweak interactions has been very successful 
when tested by experiments at and below the scale of the weak-boson masses.  
However, all available precision data concerns processes with four light 
external fermions only.  There is very little data which directly reflects the 
couplings of electroweak bosons amongst themselves, and the symmetry-breaking 
sector remains wholly uninvestigated.

Studies of the process \eeww\/ will provide important data concerning both 
non-Abelian gauge-boson couplings and the Higgs sector.   A convenient  
form-factor-based analysis of this process is indisposable if we wish to discuss 
the search for new physics effects in an efficient manner.  The utility of this 
approach has been demonstrated in Ref.~\cite{hpzh87}, where seven tensors, each 
with a scalar form-factor coefficient, were introduced to describe new physics 
in the s-channel.  However, one should expect that realistic models may induce 
corrections in the t-channel and box graphs as well.  Furthermore, it is 
desirable to have a framework which allows both standard radiative 
corrections\cite{bdsbbk88andfjz89}
and nonstandard contributions to be straightforwardly combined.  In 
Section~\ref{sec-formfactor} we demonstrate how a total of nine tensors may be 
used to obtain the most general amplitudes.  While Section~\ref{sec-formfactor} 
concentrates on the kinematics of \eeww\/ amplitudes, Section~\ref{sec-ff} 
concentrates on the dynamical structure of each form factor at the one-loop 
level in the SM supplemented by small nonstandard contributions.

We then describe deviations from the SM via effective-Lagrangian techniques.  In 
general one constructs an effective Lagrangian by adding to the SM Lagrangian 
terms which describe the new physics.  These new terms will be constructed, 
subject to the various assumptions of the extended theory,  from the fields of 
the SM and derivatives thereof.  We will everywhere assume that the full theory 
is invariant under ${\rm SU(2)}_L\times {\rm U(1)}_Y$ spontaneously broken to 
${\rm U(1)}_{\rm em}$.  Furthermore, we will assume that the couplings of the 
new physics to the light fermions are suppressed, hence fermionic fields shall 
not be employed in the construction of effective operators.  Because the 
existence or nonexistence of the Higgs boson has not yet been established, its 
inclusion or exclusion is open to debate.  We therefore consider both scenarios 
by discussing the linear and the nonlinear realizations of the symmetry-breaking 
sector.  Wherever possible we present our results in a fashion which facilitates 
comparison of the two scenarios.

In Section~\ref{sec-linear} we present an effective Lagrangian with a linearly 
realized Higgs sector which may be written as the sum of the SM Lagrangian plus 
operators of energy-dimension greater than four.  At the energy-dimension-six 
level we present a complete set of such operators, and we discuss the couplings 
affected by each operator.  In Section~\ref{sec-nonlinear} we describe the 
construction of the electroweak chiral Lagrangian.  A complete set of 
operators through energy-dimension-four are presented.  Each operator in the 
nonlinear representation is paired with its counterpart in the linear 
representation from which it may be obtained in the limit where the mass of the 
Higgs boson is taken to infinity.  We discuss the electroweak gauge-boson 
couplings which are affected by each nonlinear operator.

In Section~\ref{sec-eeff} we show how a subset of the operators, in eitherB 
realization of the symmetry-breaking sector, may be constrained by current data
from the LEP/SLC and low-energy experiments.  At low-energies three operators 
in the nonlinear representation are tightly constrained.  Four operators in the 
linear representation may be constrained, albeit somewhat less stringently, by 
the low-energy data; the constraints on these four are very much improved 
through the study of \eeff\/ at higher energies.

In Section~\ref{sec-corrections} we return to the process \eeww,\/ for which we
calculate the form-factors in the linear and in the nonlinear representation.
In either representation seven operators contribute to \eeww.\/  We also 
review the standard parameterization of the most general $WW\gamma$/$WWZ$
vertex.

In Section~\ref{sec-numerical} we present a numerical study of the process 
\eeww\/ including nonstandard effects.  In Section~\ref{sec-discussion} we 
discuss the numerical results and how they may be 
combined with constraints from the low-energy experiments, $Z$-pole data and 
further measurements of four-fermion observables at higher energies.  Finally, 
in Section~\ref{sec-conclusions} we present our conclusions.


\section{A form-factor-based analysis of \eeww}\label{sec-formfactor}

\begin{figure}[htb]
\begin{center}
\leavevmode\psfig{file=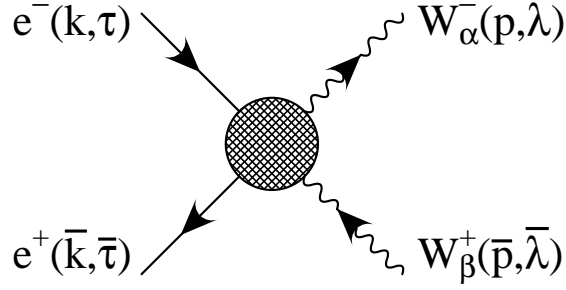,angle=0,height=4cm,silent=0}
\end{center}
\caption{The process $e^-e^+ \rightarrow W^-W^+$\/ with momentum and helicity 
assignments.The momenta $k$ and $\overline{k}$ are incoming, but $p$ and 
$\overline{p}$ are outgoing.  The arrows along the $W$-boson lines indicate 
the flow of negative electronic charge.}
\label{fig-eeww-blob} 
\end{figure}

The process $e^-(k,\tau) + e^+(\overline{k},\overline{\tau}) \rightarrow
W^-(p,\lambda) + W^+(\overline{p},\overline{\lambda})$ is depicted in 
Fig.\ref{fig-eeww-blob}.
The four-momenta of the $e^-$, $e^+$, $W^-$ and $W^+$ are $k$, $\overline{k}$,
$p$ and $\overline{p}$ respectively.  The helicity of the $e^-$ ($e^+$) is 
given by $\frac{1}{2}\tau$ ($\frac{1}{2}\overline{\tau}$), and $\lambda$
($\overline{\lambda}$) is the helicity of the $W^-$ ($W^+$).  In the limit of
massless electrons only $\overline{\tau} = -\tau$ amplitudes survive, and the 
most general amplitude for this process may be written as
\begin{equation}
\label{general-form}
{\cal M}(k,\bar{k},\tau;p,\bar{p},\lambda,\bar{\lambda})
= \sum_{i=1}^{9} F_{i,\tau}(s,t)\, j_\mu(k,\bar{k},\tau) T_i^{\mu\alpha\beta} 
\epsilon_\alpha(p,\lambda)^\ast \epsilon_\beta(\bar{p},\bar{\lambda})^\ast  \;,
\end{equation}
where all dynamical information is contained in the scalar form-factors 
$F_{i,\tau}(s,t)$ with $s = (k + \overline{k})^2$ and $t = (k - p)^2$.  The 
other factors in Eqn.~(\ref{general-form}) are of a purely kinematical nature; 
$\epsilon_\alpha(p,\lambda)^\ast$ and 
$\epsilon_\beta(\bar{p},\bar{\lambda})^\ast$ are the polarization vectors for 
the $W^-$ and $W^+$ bosons respectively, and $j_\mu(k,\bar{k},\tau)$, given by   
\begin{equation}
\label{fcurrent}
j_\mu(k,\bar{k},\tau) = \bar{v}(\bar{k},-\tau) \gamma_\mu u(k,\tau) \;,
\end{equation}
is the fermion current for massless electrons.

The tensors $T_i^{\mu\alpha\beta}$ may be chosen as
\begin{subequations}
\begin{eqnarray}
\label{ff1}
T_1^{\mu\alpha\beta} & = &  P^\mu g^{\alpha\beta} \;,\\
\label{ff2}
T_2^{\mu\alpha\beta} & = & \frac{-1}{m_W^2} P^\mu q^\alpha q^\beta \;,\\
\label{ff3}
T_3^{\mu\alpha\beta} & = & q^\alpha g^{\mu\beta} - q^\beta g^{\alpha\mu} \;,\\
\label{ff4}
T_4^{\mu\alpha\beta} & = & 
               i \Big( q^\alpha g^{\mu\beta} + q^\beta g^{\alpha\mu} \Big) \;,\\
\label{ff5}
T_5^{\mu\alpha\beta} & = & i \epsilon^{\mu\alpha\beta\rho} P_\rho \;,\\
\label{ff6}
T_6^{\mu\alpha\beta} & = & - \epsilon^{\mu\alpha\beta\rho} q_\rho \;,\\
\label{ff7}
T_7^{\mu\alpha\beta} & = & \frac{-1}{m_W^2} P^\mu 
                    \epsilon^{\alpha\beta\rho\sigma} q_\rho P_\sigma \;,\\
\label{ff8}
T_8^{\mu\alpha\beta} & = &  K^\beta g^{\alpha\mu} + K^\alpha g^{\mu\beta}\;,\\
\label{ff9}
T_9^{\mu\alpha\beta} & = &  \frac{i}{m_W^2} 
                            \Big( K^\alpha \epsilon^{\beta\mu\rho\sigma} 
             + K^\beta \epsilon^{\alpha\mu\rho\sigma} \Big) q_\rho P_\sigma\;,
\end{eqnarray}
\label{ff1thru9}
\end{subequations}
where $P = p-\bar{p}$, $q = k+\bar{k} = p+\bar{p}$, $K = k-\bar{k}$ and 
$\epsilon_{0123} = -\epsilon^{0123} = +1$.  The properties of the associated 
form factors $F_{i,\tau}(s,t)$ under the discrete transformations of charge 
conjugation ($C$), parity inversion ($P$) and the combined transformation $CP$ 
are summarised in Table~\ref{table-c-p-cp}.
\begin{table}[htb]
\begin{tabular}{|c||c|c|c|c|c|c|c|c|c|}
&\hspace*{0.3cm}$F_1$\hspace*{0.3cm} &\hspace*{0.3cm}$F_2$\hspace*{0.3cm} 
&\hspace*{0.3cm}$F_3$\hspace*{0.3cm} &\hspace*{0.3cm}$F_4$\hspace*{0.3cm} 
&\hspace*{0.3cm}$F_5$\hspace*{0.3cm} &\hspace*{0.3cm}$F_6$\hspace*{0.3cm} 
&\hspace*{0.3cm}$F_7$\hspace*{0.3cm} &\hspace*{0.3cm}$F_8$\hspace*{0.3cm} 
&\hspace*{0.3cm}$F_9$\hspace*{0.3cm} \\ 
\hline \hline
$C$  & $+$ & $+$ & $+$ & $-$ & $-$ & $+$ & $+$ & $+$ & $-$ \\ \hline
$P$  & $+$ & $+$ & $+$ & $+$ & $-$ & $-$ & $-$ & $+$ & $-$ \\ \hline
$CP$ & $+$ & $+$ & $+$ & $-$ & $+$ & $-$ & $-$ & $+$ & $+$ \\
\end{tabular}
\vspace{0.2cm}
\caption{The properties of the form factors $F_{i,\tau}(s,t)$ under the discrete 
transformations C, P and CP.}
\label{table-c-p-cp}
\end{table}

When working in the context of a 
particular model the calculation of the scalar form-factors, $F_{i,\tau}(s,t)$, 
depends upon the dynamics particular to that model as well as the level of 
precision to which the calculation is performed.  To the contrary, the 
kinematical aspects are completely general.  Therefore, it is practical to 
choose a convenient frame and to tabulate 
\begin{equation}\label{amplets}
j_\mu(k,\overline{k},\tau)\, T_i^{\mu\alpha\beta} 
\epsilon_\alpha(p,\lambda)^\ast  \epsilon_\beta(\bar{p},\bar{\lambda})^\ast  = 
\tau\sqrt{2}s\,
\widehat{T}_{i,\tau}(k,\overline{k},\tau;p,\overline{p},\lambda,\overline{\lambda})
\,d^{J_0}_{\tau,\,\Delta\lambda} 
\;,
\end{equation}
for $i = 1 \cdots 9$.   On the right-hand side of the equation an overall factor 
is extracted as well as the appropriate d-functions\cite{pdg96,rose57}, 
$d^{J_0}_{\tau,\Delta\lambda}$, where $\frac{1}{2}\tau$ is the electron helicity,
 $\Delta\lambda = \lambda-\overline{\lambda}$ and $J_0$ is the angular momentum 
of the first partial wave which contributes.  Those d-functions which are 
relevant to the current discussion are summarized in Table~\ref{table-dfunc}.
\begin{table}[tbhp]
\begin{tabular}{c}
$d^1_{\tau,0} = -\tau\sqrt{\frac{1}{2}}\sin\theta$,\hspace{.4cm} 
$d^1_{\tau,\pm 1} = \frac{1}{2} (1\pm\tau\cos\theta)$ \\
$d^2_{\tau,0} = -\tau\sqrt{\frac{3}{2}}\sin\theta \cos\theta$,\hspace{.4cm} 
$d^2_{\tau,\pm 1} = \frac{1}{2} (1\pm\tau\cos\theta) (2\cos\theta \mp \tau)$,
\hspace{.4cm} $d^2_{\tau,\pm 2} = \pm \frac{1}{2} 
(1\pm\tau\cos\theta)\sin\theta$ \\
\end{tabular}
\caption{A list of the d-functions which are used in 
Eqn.~(\protect\ref{amplets}), Table~\protect\ref{table-ampets1thru7} and 
Table~\protect\ref{table-ampets8thru9}.}
\label{table-dfunc}
\end{table}

We choose the $e^+e^-$-collision center of momentum (CM) frame with the outgoing 
$W$-boson momentum vectors along the z-axis.  The angle $\Theta$ is measured 
between the momentum vectors of the electron and the $W^-$ boson.  Then
\begin{subequations}
\label{w-z-frame}
\begin{eqnarray}
q^\mu & = & \sqrt{s} \Big( 1 , 0 , 0 , 0 \Big) \;, \\
P^\mu & = & \sqrt{s} \Big( 0 , 0 , 0 , \beta  \Big) \;, \\
K^\mu & = & \sqrt{s} \Big( 0, -\sin\Theta , 0 , \cos\Theta \Big) \;,
\end{eqnarray}
\end{subequations}
and,
in the notation of Ref.~\cite{hz86}, the fermion current and the polarization 
vectors become
\begin{subequations}
\begin{eqnarray}
j^\mu(k,\overline{k},\tau) & = & 
	\sqrt{s} \Big( 0 , -\cos\Theta , -i\tau , -\sin\Theta \Big)\;,\\
\epsilon^{\mu} (p,\pm)^\ast & = & 
	\sqrt{\frac{1}{2}} \Big( 0 , \mp 1 , i , 0 \Big) \;, \\
\epsilon^{\mu} (p,0)^\ast & = & \gamma \Big( \beta , 0 , 0 , 1 \Big) \;, \\
\epsilon^{\mu} (\overline{p},\pm)^\ast & = & 
	\sqrt{\frac{1}{2}} \Big( 0 , \mp 1 , -i , 0 \Big) \;, \\
\epsilon^{\mu} (\overline{p},0)^\ast & = & \gamma \Big( \beta , 0 , 0 , 
        -1 \Big) \;,
\end{eqnarray}
\end{subequations}
with
\begin{equation}
\beta = \sqrt{1-m_W^2/E_W^2}\;,\makebox[.5cm]{} \gamma = E_W/m_W
\;,\makebox[.5cm]{}E_W = \sqrt{s}/2  \;.
\end{equation}
The explicit form of the $\widehat{T}_{i,\tau}$ in this frame are 
summarised in Table~\ref{table-ampets1thru7} for $i = 1, \cdots, 7$, and in  
Table~\ref{table-ampets8thru9} for $i = 8,9$.  Note that the results of these 
two tables are valid in any CM frame obtained from the frame of 
Eqn.~(\ref{w-z-frame}) by a simple rotation.
\begin{table}[tbhp]
\begin{tabular}{|c|c|c||c|c|c|c|c|c|c|}
\hspace*{.2cm}$\Delta\lambda$\hspace*{.2cm} & 
\hspace*{.2cm}$\lambda\overline{\lambda}$\hspace*{.2cm} & 
\hspace*{.2cm}$d^{J_0}_{\tau,\Delta\lambda}$\hspace*{.2cm} &
  \makebox[1.2cm]{$\widehat{T}_1$} & \makebox[1.2cm]{$\widehat{T}_2$} 
& \makebox[1.2cm]{$\widehat{T}_3$} & \makebox[1.2cm]{$\widehat{T}_4$} 
& \makebox[1.2cm]{$\widehat{T}_5$} & \makebox[1.2cm]{$\widehat{T}_6$} 
& \makebox[1.2cm]{$\widehat{T}_7$} 
\\ \hline \hline
$0$  & $00$ & $d^1_{\tau,0}$&
	      $-\gamma^2\beta (1 + \beta^2)$&
              $4\beta^3\gamma^4$&
              $2\gamma^2\beta$ &&&& \\ \hline 
$0$  & $++$ & $d^1_{\tau,0}$ &
	      $ \beta $ &&&&&
	      $+i$ &
  	      $+4 i \beta^2 \gamma^2$ \\ \hline 
$0$  & $--$ & $d^1_{\tau,0}$ &
	      $ \beta $ &&&&&
	      $- i$ &
  	      $- 4 i \beta^2 \gamma^2$ \\ \hline 
$+1$ & $+0$ & $d^1_{\tau,1}$&&&
	       $\gamma\beta$&
	       $-i\gamma\beta$&
               $+\gamma\beta^2$&
               $+i\gamma$& \\ \hline 
$+1$ & $0-$ & $d^1_{\tau,1}$&&&
	       $\gamma\beta $& 
	       $+i\gamma\beta $&
	       $+\gamma\beta^2 $&
	       $-i\gamma$&  \\ \hline
$-1$ & $0+$ & $d^1_{\tau,-1}$&&& 
	       $\gamma\beta$& 
	       $+i\gamma\beta$& 
	       $- \gamma\beta^2$& 
	       $+i\gamma$&   \\ \hline
$-1$ & $-0$ & $d^1_{\tau,-1}$&&&
	       $\gamma\beta$&  
	       $- i\gamma\beta$&  
  	       $- \gamma\beta^2$& 
	       $-i\gamma$& \\ 
\end{tabular}
\caption{The $\widehat{T}_{i,\tau}(k,\overline{k},\tau;p,\overline{p},
\lambda,\overline{\lambda})$, $i = 1 \cdots 7$ evaluated in the CM frame.  
(For $i = 8, 9$ see Table~\protect\ref{table-ampets8thru9}.) In 
Eqn.~(\protect\ref{amplets}) each $\widehat{T}_{i,\tau}$ is associated with a 
d-function; for each $i$ the appropriate d-function is listed in the third 
column. In column~1, $\Delta\lambda= \lambda - \overline{\lambda}$.  Only 
nonzero results are presented.  For $i = 1 \cdots 7$ there are no contributions 
to $\Delta\lambda = 2$ amplitudes.}
\label{table-ampets1thru7}
\end{table}
\begin{table}[bth]
\begin{center}
\begin{minipage}{12cm}
\begin{tabular}{|c|c||c|c|}
\makebox[1cm]{$\Delta\lambda$} & 
\makebox[1.5cm]{$\lambda\overline{\lambda}$}& 
\makebox[3cm]{$\widehat{T}_8\,d^{J_0}_{\tau,\Delta\lambda}$}&
\makebox[5.5cm]{$\widehat{T}_9\, d^{J_0}_{\tau,\Delta\lambda}$} 
\\ \hline \hline
$0$  & $00$ &$-\sqrt{\frac{4}{3}}\gamma^2 d^2_{\tau,0}$
            & \\ \hline
$0$  & $++$ &$-\sqrt{\frac{1}{3}}d^2_{\tau,0}$
            &$- 4\gamma^2 \beta \tau d^1_{\tau,0}$ \\ \hline
$0$  & $--$ &$-\sqrt{\frac{1}{3}} d^2_{\tau,0}$
            &$- 4\gamma^2 \beta \tau d^1_{\tau,0}$ \\ \hline
$+1$ & $+0$ &$-\gamma d^2_{\tau,1}$
            &$-2 \gamma^3\beta
              (d^2_{\tau,1} + \tau d^1_{\tau,1})$ \\ \hline 
$+1$ & $0-$ &$-\gamma d^2_{\tau,1}$
            &$-2\gamma^3 \beta
              (d^2_{\tau,1} + \tau d^1_{\tau,1})$ \\ \hline
$-1$ & $0+$ &$-\gamma d^2_{\tau,-1}$
            &$2\gamma^3 \beta
              (d^2_{\tau,-1} - \tau d^1_{\tau,-1})$ \\ \hline
$-1$ & $-0$ &$-\gamma d^2_{\tau,-1}$
            &$ 2\gamma^3 \beta
              (d^2_{\tau,-1} - \tau d^1_{\tau,-1})$ \\ \hline 
$+2$ & $+-$ &$-\sqrt{2} d^2_{\tau,2}$
            &$-4\sqrt{2} \gamma^2\beta d^2_{\tau,2}$ \\ \hline 
$-2$ & $-+$ &$-\sqrt{2} d^2_{\tau,-2}$
            &$4\sqrt{2} \gamma^2\beta d^2_{\tau,-2}$ \\  
\end{tabular}
\vspace{.2cm}
\end{minipage}
\caption{The $\widehat{T}_{i,\tau}(k,\overline{k},\tau;p,\overline{p},
\lambda,\overline{\lambda})$, $i = 8,9$ evaluated in the CM frame.  
(For $i = 1\cdots 7$ see Table~\protect\ref{table-ampets1thru7}.)
Each $\widehat{T}_{i,\tau}$ is explicitly multiplied by the appropriate 
d-function or linear combination of d-functions. In column~1, 
$\Delta\lambda= \lambda - \overline{\lambda}$.  Only nonzero results are 
entered in the table.}
\label{table-ampets8thru9}
\end{center}
\vspace{-0.5cm}
\end{table}

Seven of the tensors, Eqn.~(\ref{ff1})-Eqn.~(\ref{ff7}), follow the notation of
Ref.~\cite{hpzh87}, where the primary emphasis was the discussion of nonstandard 
$WW\gamma$ and $WWZ$ vertices which respect Lorentz invariance and 
electromagnetic gauge invariance, but not SU(2)$\times$U(1) gauge invariance.
Under these assumptions the most general $WWV$ vertex ($V$ = $\gamma$ or $Z$) 
may be written
\begin{equation}\label{wwv-vertex}
\Gamma_V^{\mu\alpha\beta}(q,p,\overline{p})
 = \sum_{i=1}^{7} f_{i}^V(s) T_i^{\mu\alpha\beta}\;,
\end{equation}
where the $f_{i}^V$ are the form factors of Ref.~\cite{hpzh87}.  The two tensors 
which are new, Eqn.~(\ref{ff8})-Eqn.~(\ref{ff9}), are necessary to include all 
possible effects, including t-channel and box corrections.


\section{Calculation of the form-factors}\label{sec-ff}

In this section we write the scalar form-factors, $F_{i,\tau}(s,t)$, of 
Eqn.~(\ref{general-form}) in a form which is valid at the one-loop order for 
completely general corrections in the two- and three-point functions.  For 
higher-order effects which include fermionic vertices or the self-energy 
corrections for fermions, only the SM corrections are explicitly included. We 
find
\begin{eqnarray}
\nonumber
\lefteqn{F_{i,\tau}(s,t) = \frac{1}{s}\Bigg\{\bigg(
	Q\Big[\baresq(s) + \hatesq \Gamma_1^{\,e}(s)
     	- i \hatesq \Delta_{\gamma\gamma}(s) \Big] 
	+ I_3 \hatesq \overline{\Gamma}_2^{\,e}(s)
	\bigg)f_i^{\gamma\,(0)} 
	+ Q \hatesq f_i^{\gamma\,(1)}(s) \Bigg\} } 
\\ \nonumber && \makebox[0cm]{} 
	+ \frac{1}{s-m_Z^2 + is\frac{\Gamma_Z}{m_Z}}
	\Bigg\{ \bigg(\big(I_3-\hatssq Q\big)\hatcsq
	\Big[ \bargzsq(s) + \hatgzsq \Gamma_1^{\,e}(s)	- 
	i \hatgzsq \Delta_{ZZ}(s) \Big]
	+ I_3 \hat{c}^4\hatgzsq \overline{\Gamma}_2^{\,e}(s) \bigg) 
	f_i^{Z\,(0)}  
\\ \nonumber && \makebox[0cm]{}
	+  \big(I_3-\hatssq Q\big)\hatcsq\hatgzsq f_i^{Z\,(1)}(s) 
	- \hatgzsq\Big[ Q \hatcsq f_i^{Z\,(0)} + \big(I_3-\hatssq Q\big)  
	f_i^{\gamma\,(0)} \Big] \Big( \barssq(s) - \hatssq 
	+ i \Delta_{\gamma Z}(s)\Big) \Bigg\}
\\ && \makebox[0cm]{} 
	+ \frac{1}{2t} I_3 \hatgsq \Big( 1 + \Gamma^{\,e\nu}(t) 
	+ \overline{\Gamma}^{\,e\nu}(t) \Big)f_i^{t\,(0)} + B_{i,\tau}(s,t)\;,
\label{big-ff}
\end{eqnarray}
where the hatted couplings, $\hate = \hatg\hats = \hatg^\prime\hatc = 
\hatgz\hatc\hats$, are the $\overline{\rm MS}$ couplings with a short-hand 
notation $\hatssq = 1 - \hatcsq = 
\sin^2\hat{\theta}_W(\mu)_{\overline{\rm MS}}$.   $I_3$ and $Q$ refer to the 
SU(2)-isospin and electromagnetic-charge quantum numbers of the electron, \ie\/
$Q=-1$, and $I_3 = -1/2$ ($I_3 = 0$) for a left-handed (right-handed) electron.
Gauge-boson propagator corrections are contained in the form factors
$\baresq(s)$, $\bargzsq(s)$, $\barssq(s)$, $\Delta_{\gamma\gamma}(s)$, 
$\Delta_{\gamma Z}(s)$ and $\Delta_{ZZ}(s)$; $\Gamma_1^{\,e}(s)$ and 
$\overline{\Gamma}_2^{\,e}(s)$ contain corrections to the $eeV$ vertex
and $e^\pm$ self-energy corrections\cite{hhkm94}.

\begin{figure}[htb]
\begin{center}
\leavevmode\psfig{file=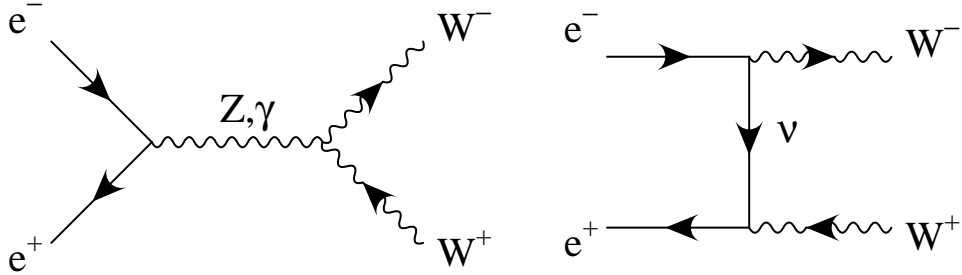,angle=0,height=4cm,silent=0}
\end{center}
\caption{The SM Feynman graphs for the process $e^-e^+ \rightarrow W^-W^+$.\/ 
Momentum and helicity assignments coincide with those of 
Fig.~\protect\ref{fig-eeww-blob}.}
\label{fig-eeww-sm} 
\end{figure}

In Eqn.~(\ref{big-ff}) we introduce additional form-factors through
\begin{equation}
\label{small-ff}
f_i^X(s) = f_i^{X\,(0)} 
+  f_i^{X\,(1)}(s)\;,
\end{equation}
where $X=Z,\gamma,t$. At the tree level, $f_i^{\gamma\,(0)}$ ($f_i^{Z\,(0)}$), 
for $i = 1, \cdots, 7$, corresponds to the SM contribution to the s-channel 
exchange of a photon ($Z$ boson).  See Fig.~\ref{fig-eeww-sm}.  The values of 
the $f_i^{V\,(0)}$ may be 
obtained via the expansion of the tree-level $WWV$ vertex according to 
Eqn.~(\ref{wwv-vertex}).  At higher orders there may be corrections directly to
the $WWV$ vertex, which are contained in $f_i^{V\,(1)}$.  Associated 
self-energy corrections for the external $W$ bosons are also included in 
$f_i^{V\,(1)}$.

In a similar fashion the t-channel contribution to the tree-level amplitude 
may be expanded to obtain $f_i^{t\,(0)}$ for $i = 1, \cdots, 9$.  We do not 
introduce an $f_i^{t\,(1)}$ term. Such a term would correspond to corrections
to the $We\nu$ vertex beyond the SM;  as stated above, nonstandard couplings 
to the external fermions are not explicitly considered.  The only nonstandard 
corrections which enter via the neutrino-exchange diagram are the $W$-boson 
self-energy corrections which are included in the $e^-\nu W^-$ vertex-correction 
factor,  $\Gamma^{\,e\nu}$, and in the $e^+\overline{\nu} W^+$ vertex-correction 
factor, $\overline{\Gamma}^{\,e\nu}$.  

The SM tree-level values for the $f^{X\,(0)}_i$ are shown in 
Table~\ref{table-smff}.
\begin{table}[htb]
\begin{tabular}{|c||c|c|c|c|c|c|c|c|c|}
&\hspace*{0.3cm}$i=1$\hspace*{0.3cm} &\hspace*{0.3cm}$i=2$\hspace*{0.3cm} 
&\hspace*{0.3cm}$i=3$\hspace*{0.3cm} &\hspace*{0.3cm}$i=4$\hspace*{0.3cm} 
&\hspace*{0.3cm}$i=5$\hspace*{0.3cm} &\hspace*{0.3cm}$i=6$\hspace*{0.3cm} 
&\hspace*{0.3cm}$i=7$\hspace*{0.3cm} &\hspace*{0.3cm}$i=8$\hspace*{0.3cm} 
&\hspace*{0.3cm}$i=9$\hspace*{0.3cm} \\ 
\hline \hline
$f_{i}^{\gamma\,(0)}$ & 1 & 0 & 2 & 0 & 0 & 0 & 0 & 0 & 0 \\ \hline
$f_{i}^{Z\,(0)}$      & 1 & 0 & 2 & 0 & 0 & 0 & 0 & 0 & 0 \\ \hline
$f_{i}^{t\,(0)}$      & 1 & 0 & 2 & 0 & 1 & 0 & 0 & 1 & 0 \\
\end{tabular}
\vspace{0.2cm}
\caption{Explicit values for the $f_i^{X\,(0)}$ form factors of the SM at the 
tree level.}
\label{table-smff}
\end{table}
Refering to Table~\ref{table-c-p-cp}, we see that $f_1^{\gamma\,(0)} =
f_1^{Z\,(0)} = f_1^{t\,(0)} = 1$ all contribute to the $C$-even $P$-even 
form-factor $F_{1,\tau}(s,t)$.  Similarly $f_3^{\gamma\,(0)} = f_3^{Z\,(0)} = 
f_3^{t\,(0)} = 2$ all contribute to the $C$-even $P$-even form-factor 
$F_{3,\tau}(s,t)$.  Parity violation in the SM tree-level amplitudes enters 
through the $C$-odd $P$-odd form-factor $F_{5,\tau}(s,t)$; that parity 
violation appears only via the t-channel Feynman graph of Fig.~\ref{fig-eeww-sm}
is apparent from the values $f_5^{\gamma\,(0)} = f_5^{Z\,(0)} = 0$,
$f_5^{t\,(0)} = 1$.  Finally, spin-greater-than-one contributions are manifest 
through the contribution of $f_8^{t\,(0)}=1$ to the $C$-even $P$-even form-factor
$F_{8,\tau}(s,t)$.  The regular pattern that appears in 
Table~\ref{table-smff} is extremely important, as will be discussed at the end
of this section in the context of tree-level perturbative unitarity.  While
the SM employs neither the $C$-even $P$-even form-factor $F_{2,\tau}(s,t)$ nor
the $C$-odd $P$-odd form-factor $F_{9,\tau}(s,t)$ at the tree level, at the
one-loop level they attain nonzero values\cite{bdsbbk88andfjz89}.  
$F_{9,\tau}(s,t)$ is generated solely through box corrections.

The barred charges include the real parts of the gauge-boson two-point-functions
\cite{hhkm94};
\begin{subequations}
\label{def-barcharges}
\begin{eqnarray}
\label{def-ebar}
\baresq(q^2) & = & \hatesq\bigg[ 1 -  {\rm Re} 
	\overline{\Pi}^{\gamma\gamma}_{T,\gamma}(q^2)\bigg]\;,
\\ \label{def-sbar}
\barssq(q^2) & = & \hatssq\bigg[ 1 + 
      \frac{\hatc}{\hats} {\rm Re} 
	\overline{\Pi}^{Z\gamma}_{T,\gamma}(q^2)\bigg]\;,
\\ \label{def-gzbar}
\bargzsq(q^2) & = & \hatgzsq\bigg[ 1 -  {\rm Re} 
		\overline{\Pi}^{ZZ}_{T,Z}(q^2)\bigg]\;,
\\ \label{def-gwbar}
\bargwsq(q^2) & = & \hatgsq\bigg[ 1 -  {\rm Re} 
		\overline{\Pi}^{WW}_{T,W}(q^2)\bigg]\;.
\end{eqnarray}
\end{subequations}
While $\baresq$, $\barssq$ and $\bargzsq$ are employed explicitly in 
Eqn.~(\ref{big-ff}), $\bargwsq$ enters only through the $W$-boson 
wave-function-renormalization factor  as discussed below.  The 
$\Delta_{VV^\prime}$, which are the imaginary parts of the 
two-point-function corrections, are given by
\begin{subequations}
\label{imaginary-parts}
\begin{eqnarray}
\label{deltagg}
\Delta_{\gamma\gamma}(q^2) & = & {\rm Im} 
  \overline{\Pi}^{\gamma\gamma}_{T,\gamma}(q^2)\;,
\\
\label{deltagz}
\Delta_{\gamma Z}(q^2) & = & \hats\hatc\, {\rm Im}
  \overline{\Pi}^{\gamma Z}_{T,\gamma}(q^2)\;,
\\
\label{deltazz}
\Delta_{ZZ}(q^2) & = & {\rm Im}\overline{\Pi}^{ZZ}_{T,Z}(q^2) -
  \frac{{\rm Im}\overline{\Pi}^{ZZ}_{T}(m_Z^2)}{m_Z^2}\;.
\end{eqnarray}
\end{subequations}
Here
\begin{eqnarray}\label{defn-pitv}
\overline{\Pi}^{AB}_{T,V}(q^2) = {\overline{\Pi}^{AB}_T(q^2) - 
\overline{\Pi}^{AB}_T(m^2_V)\over q^2 -
m^2_V}\;,
\end{eqnarray}
where $m_V$ denotes the physical mass of the gauge boson $V$.  The subscript 
`T' indicates the use of the transverse component of gauge-boson 
two-point-function; the longitudinal component makes no contribution when 
coupled to an external massless-fermion current.  We employ the LEP convention 
for the $Z$-boson mass and running width\cite{sir91} which accounts for the 
additional contribution to $\Delta_{ZZ}$ in Eqn.~(\ref{deltazz}).
The pinch-term contributions 
\cite{kl89,cp89andpap90,ds92anddks93,pinch} have been removed 
from the vertex-correction terms, \ie\/ $\overline{\Gamma}^{\,e}_2(s)$ 
(and also $f_1^{V\,(1)}(s)$ and $f_3^{V\,(1)}(s))$, but instead have been 
absorbed into the barred effective charges\cite{hhkm94}.  This standard 
procedure renders 
the effective charges gauge invariant and allows us to use them universally in 
both the four-fermion and \eeww\/ amplitudes.

The $\Gamma_1^{\,e}$ and $\overline{\Gamma}_2^{\,e}$ terms contain the 
corrections to the $ee\gamma$ and $eeZ$ vertices as well as the 
associated self-energy corrections of the external electrons(positrons).  
The $e^- \nu W^-$ ($e^+ \overline{\nu} W^+$) vertex corrections are combined 
with the electron (positron) and $W^-$ ($W^+$) wave-function renormalization 
factors and one half (one half) of the internal neutrino self-energy 
corrections to produce finite 
form factors $\Gamma^{\,e\nu}$ ($\overline{\Gamma}^{\,e\nu}$).  The final term 
in Eqn.~(\ref{big-ff}), $B_{i,\tau}(s,t)$, includes all box-type 
corrections.

Finally, we conclude this section with a discussion of the \eeww\/ amplitudes
at the tree level in the SM.  Notice that, in Table~\ref{table-ampets1thru7} and 
Table~\ref{table-ampets8thru9}, the contributions where either one or two $W$ 
bosons are longitudinally polarized grow at high energies as a power of the 
kinematical variable $\gamma = E_W/m_W$. If a perturbative description remains 
valid at high energies, then the tree-level unitarity of the amplitudes demands 
that these large contributions cancel among the various terms of 
Eqn.~(\ref{general-form}) and Eqn.~(\ref{big-ff}).  These cancellations are 
straightforwardly displayed in the current formalism.

Combining Eqns.~(\ref{general-form}), (\ref{big-ff}) and (\ref{amplets}), then 
taking the limit $\beta \rightarrow 1$, $\gamma \rightarrow \infty$, the leading 
contributions to the amplitudes may be expressed as
\begin{equation}\label{amp-limit}
{\cal M}(\tau;\lambda,\bar{\lambda})
= \sum_{i=1}^{9} \tau\sqrt{2} \Bigg\{ \hatesq Q f^{\gamma\,(0)}_i
	+ \Big(\hatgsq I_3 - \hatesq Q \Big) f^{Z\,(0)}_i
	- \frac{\hatgsq I_3}{1-\cos\Theta}f^{t\,(0)}_i     \Bigg\}
	\widehat{T}_i\,d^{J_0}_{\tau\,\Delta\lambda}\;.
\end{equation}
On the left-hand side momentum arguments have been suppressed for brevity.
First, consider the amplitude for $\lambda\bar{\lambda} = 00$.  From 
Table~\ref{table-smff} we see that $\widehat{T}_1$, $\widehat{T}_3$ and 
$\widehat{T}_8$ contribute to this amplitude.  In Eqn.~(\ref{amp-limit}) it is 
immediately apparent that the nonzero terms proportional to $\hatesq Q$ vanish 
within each $F_{i,\tau}(s,t)$ because $f^{\gamma\,(0)}_i = f^{Z\,(0)}_i$ for 
both $i=1$ and $i=3$.  For left-handed fermion currents the $\hatgsq I_3$ terms 
also play a role.  However, in this case the cancellations only take place upon 
summation over the various form-factors.  Using $d^2_{\tau,0}=\sqrt{3}\cos\Theta 
d^1_{\tau,0}$ for the treatment of the $\widehat{T}_{8,-}$ contribution one 
obtains
\begin{eqnarray}
\nonumber
{\cal M}(-;0,0) & \sim & \sqrt{2} \hatgsq I_3
	\Bigg\{ \frac{ -2\gamma^2d^1_{-,0}  + 4\gamma^2 d^1_{-,0} 
	- \sqrt{4/3}\gamma^2 d^2_{-,0} }{1-\cos\Theta} 
	- \Big( -2\gamma^2d^1_{-,0} + 4\gamma^2d^1_{-,0}\Big)\Bigg\}  \\
& \rightarrow & \sqrt{2} \hatgsq I_3
	\Bigg\{ \frac{2\gamma^2(1-\cos\Theta)}{1-\cos\Theta} 
	-2\gamma^2 \Bigg\}d^1_{-,0} \;,
\label{amp-00}
\end{eqnarray}
and we see that the terms proportional to $\gamma^2$ cancel.  

We may repeat this procedure for the $+0$ amplitude. The nonzero terms 
proportional to $\hatesq Q$ cancel within $F_3$. However, for the terms 
proportional to $\hatgsq I_3$, the cancellations only take place when the 
$i=3,5$ and $8$ contributions are summed.  Using $d^2_{\tau,\pm 1} = 
(2\cos\Theta\mp\tau)d^1_{\tau,\pm 1}$,
\begin{eqnarray}
\nonumber
{\cal M}(-;+,0) & \sim &  \sqrt{2} \hatgsq I_3
	\Bigg\{ \frac{ 2\gamma\,d^1_{-,1} + \gamma\,d^1_{-,1} 
	- \gamma\, d^2_{-,1}}{1-\cos\Theta} - 2\gamma\,d^1_{-,1} \Bigg\}\\
& \rightarrow & \sqrt{2} \hatgsq I_3
	\Bigg\{ \frac{2\gamma(1-\cos\Theta)}{1-\cos\Theta} 
	-2\gamma\Bigg\}d^1_{-,1} \;,
\label{amp-+0}
\end{eqnarray}
and here the contributions proportional to $\gamma$ cancel.  We could repeat 
this analysis for the $0-$, $0+$ and $-0$ amplitudes, again to discover that 
the terms which grow with energy cancel.  

	There is an advantage of adopting the $\overline{\rm MS}$ couplings
in the perturbative expansion of the form-factors.  In brief, through a 
judicious choice of the renormalization scale for these couplings, the above 
tree-level unitarity cancellation straightforwardly prevails in the major part 
of the corrected amplitudes.  In the numerical studies of subsequent sections 
we adopt the 
renormalization conditions $\hatesq=\baresq(s)_{\rm SM}$ and 
$\hatssq=\barssq(s)_{\rm SM}$. The nonstandard corrections described by an 
effective Lagrangian appear to violate tree-level unitarity, but one should 
recall that the effective-Lagrangian description is valid only at energies
below the threshold of the new physics.  This will be revisited in 
Section{\ref{sec-corrections}.


\section{The linear representation}\label{sec-linear}

If the scale of new physics, $\Lambda$, is large compared to the vacuum 
expectation value (vev) of the Higgs field, $v \equiv (\sqrt{2}G_F)^{-1/2} = 
246$GeV, then the effective Lagrangian may be expressed as the SM Lagrangian 
plus terms with energy dimension greater than four suppressed by inverse powers 
of $\Lambda$, \ie
\begin{equation}\label{fulllagrangian}
{\cal L}_{eff} = {\cal L}_{\rm SM} 
+  \sum_{n \geq 5} \sum_i \frac{f_i^{(n)} \, 
{\cal O}_i^{(n)}}{\Lambda^{n-4}} \;.
\end{equation}
The energy dimension of each operator is denoted by $n$, and the index $i$ sums 
over all operators of the given energy dimension.  The coefficients $f_i^{(n)}$
are free parameters, though they may be determined explicitly once the full 
theory is known.

The higher-dimensional terms are constructed from the fields of the low-energy 
theory.  In this section we assume that the low-energy theory, \ie\/ the SM, 
contains a light physical scalar Higgs particle which is the remnant of a 
complex Higgs-doublet field; the remaining three real fields of this doublet 
provide the longitudinal modes of the $W^\pm$ and $Z$ bosons.  We will refer to 
the physics described by the effective Lagrangian of Eqn.~(\ref{fulllagrangian}) 
as the `light-Higgs scenario' or the `linear realization of the 
symmetry-breaking sector'.

An exhaustive list of SU(2)$\times$U(1)-gauge-invariant energy-dimension-five 
and -six operators has been compiled in Refs.~\cite{bs83andbw86andllr86}.  As 
outlined in Section~\ref{sec-intro}, we exclude all operators which contain 
fermionic fields.
Furthermore, we only consider operators which conserve CP.  Upon restricting the 
analysis to operators not exceeding energy-dimension six we find that twelve 
operators form a basis set; all are dimension-six and separately conserve C and 
P.  In the notation of Ref.~\cite{hisz92,hisz93} they are
\begin{subequations}
\label{twelve-operators}
\begin{eqnarray}
\label{odw}
{\cal O}_{DW} & = & {\rm Tr} \bigg( \Big[ D_\mu,\hat{W}_{\nu\rho} \Big] 
 \, 
\Big[ D^\mu,\hat{W}^{\nu\rho} \Big] \bigg) \rule[-4mm]{0mm}{9mm} \; ,  \\
\label{odb}
{\cal O}_{DB} & = & -\frac{g'^2}{2}\Big(\partial_\mu
B_{\nu\rho}\Big)\Big(\partial^\mu B^{\nu\rho}\Big) \rule[-4mm]{0mm}{9mm} \; , \\
\label{obw}
{\cal O}_{BW} & = & \Phi^ \dagger \hat{B}_{\mu\nu} \hat{W}^{\mu\nu} 
\Phi \rule[-4mm]{0mm}{9mm} \; , \\
\label{ophi1}
{\cal O}_{\Phi,1} & = & \bigg[ \Big(D_\mu\Phi\Big)^\dagger \Phi\bigg] \; 
\bigg[ \Phi^\dagger \Big(D^\mu \Phi\Big)\bigg] \rule[-4mm]{0mm}{9mm} \;, \\
\label{owww}
{\cal O}_{WWW} &=& \tr \Big( \hat{W}_{\mu\nu} \hat{W}^{\nu\rho} 
\hat{W}_\rho\,^\mu \Big) \rule[-4mm]{0mm}{9mm} \;, \\
\label{oww}
{\cal O}_{WW} & = & \Phi^\dagger \hat{W}_{\mu\nu} \hat{W}^{\mu\nu} \Phi 
 \rule[-4mm]{0mm}{9mm} \;,  \\
\label{obb}
{\cal O}_{BB} & = & \Phi^\dagger \hat{B}_{\mu\nu} \hat{B}^{\mu\nu} \Phi 
 \rule[-4mm]{0mm}{9mm} \;, \\
\label{ow}
{\cal O}_W & = & \Big( D_\mu\Phi \Big)^\dagger \hat{W}^{\mu\nu} 
\Big( D_\nu\Phi \Big)  \rule[-4mm]{0mm}{9mm} \;,  \\
\label{ob}
{\cal O}_B & = & \Big( D_\mu\Phi \Big)^\dagger \hat{B}^{\mu\nu}
\Big( D_\nu\Phi \Big)  \rule[-4mm]{0mm}{9mm} \;,  \\
\label{op2}
{\cal O}_{\Phi,2} & = & \frac{1}{2} \partial_\mu \Big( \Phi^\dagger \Phi \Big)
\partial^\mu \Big( \Phi^\dagger \Phi \Big) \rule[-4mm]{0mm}{9mm} \;, \\
\label{op3}
{\cal O}_{\Phi,3} & = & \frac{1}{3} \Big( \Phi^\dagger \Phi \Big)^3 
\rule[-4mm]{0mm}{9mm} \;, \\
\label{op4}
{\cal O}_{\Phi,4} &  = &  \Big( \Phi^\dagger \Phi \Big) 
\bigg[\Big( D_\mu \Phi \Big)^\dagger \Big( D^\mu \Phi \Big)\bigg] 
\rule[-4mm]{0mm}{9mm}  \;.
\end{eqnarray}
\end{subequations}
The covariant derivative, $\dcov$, is given by
\begin{equation}
\dcov_\mu = \partial_\mu + i g T^a W_\mu^a + ig^\prime Y B_\mu \;,
\end{equation}
where $g$ is the SU(2) 
coupling with ${\rm Tr}(T^aT^b)=\frac{1}{2}\delta^{ab}$, 
$g^\prime$ is the U(1) coupling and $Y$ is the hypercharge operator.  For 
convenience when defining the normalizations of the individual operators 
we use the `hatted' field strength tensors defined according to
\begin{equation}
\Big[ D_\mu, D_\nu \Big] = \hat{W}_{\mu\nu} + \hat{B}_{\mu\nu}, \label{comu}
\end{equation}
hence
\begin{equation} \label{vmunuhat}
\hat{W}_{\mu\nu}  = i g T^a W^a_{\mu\nu} 
\makebox[1cm]{} {\rm and} \makebox[1cm]{}
\hat{B}_{\mu\nu} = i g^\prime Y B_{\mu\nu} \;. 
\end{equation} 
Combining the twelve operators of Eqn.~(\ref{twelve-operators}) with 
Eqn.~(\ref{fulllagrangian}) completes the construction of the effective 
Lagrangian in the linear representation.  

The calculation of the Feynman rules from Eqns.~(\ref{fulllagrangian})  and
(\ref{twelve-operators}) is straightforward, though tedious.  We do not present 
the Feynman rules, but in Table~\ref{table-linear} we indicate those vertices to 
which each operator contributes with an `{\cal X}' in the appropriate box.  
First, observe that four of the operators, ${\cal O}_{DW}$, ${\cal O}_{DB}$, 
${\cal O}_{BW}$ and ${\cal O}_{\Phi,1}$, contribute to gauge-boson 
two-point-functions at the tree level\cite{gw91}.  For this reason their 
respective 
coefficients, $f_{DW}$, $f_{DB}$, $f_{BW}$ and $f_{\Phi,1}$, are strongly 
constrained by LEP/SLC and low-energy data\cite{hisz92,hisz93,hms95}, and 
these constraints will be improved by the study of two-fermion final states at 
higher-energy lepton colliders\cite{hms95}.  (This will be discussed in 
greater detail in Section~\ref{sec-eeff}.)  These four operators will 
contribute to the process \eeww\/ through corrections to the charge 
form-factors, $\baresq(q^2)$, $\barssq(q^2)$, $\bargzsq(q^2)$, and 
through the $W$-boson wave-function-renormalization factor.

The operators ${\cal O}_{DW}$ and ${\cal O}_{BW}$ also make a direct 
contribution to $WW\gamma$ and $WWZ$ vertices.  Three additional operators 
contribute as well\cite{hisz92,hisz93}.  They are ${\cal O}_{WWW}$, 
${\cal O}_W$ and ${\cal O}_B$; their respective coefficients are 
$f_{WWW}$, $f_W$ and $f_B$.    

Naively one would expect contributions to $WWV$ three-point-functions and to 
gauge-boson two-point-functions from the operators ${\cal O}_{WW}$ and 
${\cal O}_{BB}$.  However, their contributions may be completely absorbed by a 
redefinition of SM fields and gauge couplings,
\begin{subequations}
\label{fbbfwwredef}
\begin{eqnarray}
\Bigg[1+\frac{2\hat{m}_W^2}{\Lambda^2}f_{WW}\Bigg]^{\frac{1}{2}} W^{\mu\nu} 
   & \longrightarrow & W^{\mu\nu}\;,\\
\Bigg[1+\frac{2\hat{m}_W^2}{\Lambda^2}f_{WW}\Bigg]^{-\frac{1}{2}} g 
   \makebox[.3cm]{}  & \longrightarrow & g \;,\\
\Bigg[1+\frac{2\hat{m}_Z^2}{\Lambda^2}\hatssq f_{BB}\Bigg]^{\frac{1}{2}} 
   B^{\mu\nu} 
   & \longrightarrow & B^{\mu\nu}\;,\\
\Bigg[1+\frac{2\hat{m}_Z^2}{\Lambda^2}\hatssq f_{BB}\Bigg]^{-\frac{1}{2}} 
   g^{\prime}
   \makebox[.1cm]{} & \longrightarrow & g^{\prime} \;,
\end{eqnarray}
\end{subequations}
leading to a null contribution.  For this reason an `{\cal O}' is used for 
these operators in Table~\ref{table-linear}.

Additionally ${\cal O}_{\Phi,4}$ contributes to the $W$- and $Z$-mass terms, 
while ${\cal O}_{\Phi,1}$ contributes to the $Z$-mass term only.  Hence
${\cal O}_{\Phi,1}$ violates the custodial symmetry\cite{ejv81}, SU(2)$_c$, 
and the $T$ 
parameter\cite{pt90andpt92} is explicitly dependent upon $f_{\Phi,1}$.  On the 
other hand, the contributions from ${\cal O}_{\Phi,4}$ exactly cancel in the 
calculation of $T$, hence it does not contribute to our analysis.

Notice that ${\cal O}_{DW}$, ${\cal O}_{WWW}$ and ${\cal O}_{W}$ contribute to 
four-gauge-boson vertices, though none contribute to a $ZZZZ$ vertex.  
Furthermore, many of the operators (\ref{twelve-operators}) do contribute to 
processes which include Higgs bosons.  For example, ${\cal O}_{WW}$ and 
${\cal O}_{BB}$ contribute to the $H\gamma\gamma$ vertex\cite{hsz93andhs94}.  
The operators ${\cal O}_{\Phi,2}$, ${\cal O}_{\Phi,3}$ and ${\cal O}_{\Phi,4}$ 
are of concern only when discussing nonstandard Higgs-boson interactions.

\begin{table}[htbp]
\begin{tabular}{|l||l|l|l|l|l|l|l|l|l|l|l|}
\makebox[1cm]{}\raisebox{8mm}{${\cal O}^{(6)}_i$} &
\rotate[l]{\makebox[2cm][c]{WW}} & \rotate[l]{\makebox[2cm][c]{ZZ}} & 
\rotate[l]{\makebox[2cm][c]{AZ}} & \rotate[l]{\makebox[2cm][c]{AA}} & 
\rotate[l]{\makebox[2cm][c]{WWZ}} & \rotate[l]{\makebox[2cm][c]{WWA}} & 
\rotate[l]{\makebox[2cm][c]{WWWW}} & \rotate[l]{\makebox[2cm][c]{WWZZ}} & 
\rotate[l]{\makebox[2cm][c]{WWZA}} & \rotate[l]{\makebox[2cm][c]{WWAA}} & 
\rotate[l]{\makebox[2cm][c]{ZZZZ}} \\ \hline \hline
\raisebox{-0.2cm}{${\cal O}_{DW}  =  
{\rm Tr} \bigg( \Big[ D_\mu,\hat{W}_{\nu\rho} \Big] \, 
\Big[ D^\mu,\hat{W}^{\nu\rho} \Big] \bigg)$  } 
&\cm &\cm &\cm &\cm &\cm &\cm &\cm 
&\cm &\cm &\cm & \makebox[0.4cm]{} \\[0.4cm] \hline
\raisebox{-0.2cm}{${\cal O}_{DB}  =  -\frac{g'^2}{2}\Big(\partial_\mu
B_{\nu\rho}\Big)\Big(\partial^\mu B^{\nu\rho}\Big)$ } &
& \cm & \cm & \cm &&&&&&& \\[0.4cm] \hline
\raisebox{-0.2cm}{${\cal O}_{BW} = \Phi^ \dagger \hat{B}_{\mu\nu} 
\hat{W}^{\mu\nu} \Phi$ } &
& \cm & \cm & \cm & \cm & \cm &&&&& \\[0.4cm] \hline
\raisebox{-0.2cm}{${\cal O}_{\Phi,1} = 
\bigg[ \Big(D_\mu\Phi\Big)^\dagger \Phi\bigg] \; 
\bigg[ \Phi^\dagger \Big(D^\mu \Phi\Big)\bigg] $} &
& \cm &&&&&&&&& \\[0.4cm] \hline
\raisebox{-0.2cm}{${\cal O}_{WWW} = 
\tr \Big( \hat{W}_{\mu\nu} \hat{W}^{\nu\rho} \hat{W}_\rho\,^\mu \Big) $} &
&&&& \cm & \cm & \cm & \cm & \cm & \cm & \\[0.4cm] \hline
\raisebox{-0.2cm}{${\cal O}_{WW}  =  \Phi^\dagger \hat{W}_{\mu\nu} 
\hat{W}^{\mu\nu} \Phi $} &
\cmo & \cmo & \cmo & \cmo & \cmo & \cmo & \cmo & \cmo & \cmo & \cmo 
& \\[0.4cm] \hline
\raisebox{-0.2cm}{${\cal O}_{BB} = 
\Phi^\dagger \hat{B}_{\mu\nu} \hat{B}^{\mu\nu} \Phi  $ } &
& \cmo & \cmo & \cmo &&&&&&& \\[0.4cm] \hline
\raisebox{-0.2cm}{$ {\cal O}_W = \Big( D_\mu\Phi \Big)^\dagger \hat{W}^{\mu\nu} 
\Big( D_\nu\Phi \Big)$ } &
&&&& \cm & \cm & \cm & \cm & \cm && \\[0.4cm] \hline
\raisebox{-0.2cm}{${\cal O}_B = \Big( D_\mu\Phi \Big)^\dagger \hat{B}^{\mu\nu}
\Big( D_\nu\Phi \Big) $ } &
&&&& \cm & \cm &&&&& \\[0.4cm] \hline
\raisebox{-0.2cm}{${\cal O}_{\Phi,2} = 
\frac{1}{2} \partial_\mu \Big( \Phi^\dagger \Phi \Big)
\partial^\mu \Big( \Phi^\dagger \Phi \Big)   $ } &
&&&&&&&&&& \\[0.4cm] \hline
\raisebox{-0.2cm}{${\cal O}_{\Phi,3}  = 
 \frac{1}{3} \Big( \Phi^\dagger \Phi \Big)^3  $ } &
&&&&&&&&&& \\[0.4cm] \hline
\raisebox{-0.2cm}{$ {\cal O}_{\Phi,4}  =  \Big( \Phi^\dagger \Phi \Big) 
\bigg[\Big( D_\mu \Phi \Big)^\dagger \Big( D^\mu \Phi \Big)\bigg]  $ } &
 \cmo & \cmo &&&&&&&&& \\[0.4cm]
\end{tabular}
\vspace{.2cm}
\caption{Energy-dimension-six operators in the linear representation of the 
Higgs mechanism.  The contribution of an operator to a particular vertex is 
denoted by an \bf `X' \rm.  In some cases an operator naively contributes to a 
vertex, yet that contribution does not lead to observable effects.  In such 
cases the \bf `X' \rm is replaced by an \bf `O' \rm.} 
\label{table-linear}
\end{table}


\section{The nonlinear realization}\label{sec-nonlinear}

The construction of the effective Lagrangian requires knowledge of the
low-energy particle spectrum.  The existence of a light Higgs boson has not 
been confirmed, and an intriguing possibility is that no such particle exists.
The scale for the new physics is then set by the scale of electroweak symmetry 
breaking, $v$.  Typically
\begin{equation}
\label{chiralscale}
\Lambda \sim 4\pi v\;.
\end{equation}

In general one should expect that the list of operators which contribute to 
the effective Lagrangian are related to those of the linear representation of
Eqn.~(\ref{fulllagrangian}), but the operators which appear at leading order 
may be quite different than those enumerated in 
Eqn.~(\ref{twelve-operators}).  This may be seen by studying the nonlinear 
representation of the Higgs doublet field;
\begin{equation}
\Phi(x) = \exp\Bigg(\frac{i\chi^i (x) \tau^i}{v}\Bigg)
\left( \begin{array}{c} 0 \\ (v+H)/\sqrt{2} \end{array} \right)\;,
\end{equation}
where $\chi^i (x)$ are the Goldstone fields, $H$ is the usual Higgs field and 
the $\tau^i$ are the Pauli matrices.  In the limit that the Higgs field is too 
massive to fluctuate the $H$ term may be dropped.  Then, in the unitary 
gauge, 
\begin{equation}
\Phi(x) = \frac{v}{\sqrt{2}}
\left( \begin{array}{c} 0 \\ 1 \end{array} \right)\;.
\end{equation}
Therefore, if one starts with an operator of energy dimension $n$ in the linear 
representation but removes $m$ Higgs fields, $H$, the residual operator may,  by 
Eqn.~(\ref{chiralscale}), contain a coefficient proportional to $v^m/\Lambda^n 
\sim 1/v^{n-m}$ for integers $n$ and $m$.  Hence operators which appear at 
higher orders in the linear representation may appear at a reduced order in the 
nonlinear representation.  Powers of $4\pi$ are absorbed into the numerical 
coefficients.  

The full Lagrangian may be written as
\begin{equation}
\label{fulllagrangiannl}
{\cal L}_{\rm eff} = {\cal L}_{\rm SM} + \sum_i {\cal L}_i + \cdots \;.
\end{equation}
In contrast to the linearly realized Lagrangian of Eqn.~(\ref{fulllagrangian}), 
the ${\cal L}_{\rm SM}$ term does not contain the physical Higgs field.  We 
adopt the notation\cite{lon80,lon81,aw93}
\begin{subequations}
\begin{eqnarray}
\label{u-general}
U & \equiv & \frac{\sqrt{2}}{v} \Big( \Phi^c , \Phi \Big) 
= \exp\bigg(\frac{2i\chi^i(x)\tau^i}{v} \bigg)\;, \\
\label{dcov-general}
\dcov_\mu U & = & \partial_\mu U + i g T^a W_\mu^a U - i g^\prime U T^3 B_\mu 
\;, \\
\label{tee-general}
T & \equiv & 2 U T^3 U^\dagger \;, \\
\label{v-general}
V_\mu & \equiv & ( \dcov_\mu U ) U^\dagger \;.
\end{eqnarray}
\end{subequations}
Here $\Phi^c = i\tau^2\Phi^\ast$ denotes the charge-conjugate Higgs doublet 
field, and $T^a = \tau^a/2$ are the generators of the SU(2) algebra.
In the unitary gauge these expressions become,
\begin{subequations}
\begin{eqnarray}
\label{u-unitary}
U & = & \id \;, \\
\label{dcov-unitary}
\dcov_\mu U & = &  i g T^a W_\mu^a - i g^\prime T^3 B_\mu \;, \\
\label{tee-unitary}
T & = & 2 T^3  \;, \\
\label{v-unitary}
V_\mu & = &  \dcov_\mu U \;.
\end{eqnarray}
\end{subequations}
The custodial SU(2)$_{\rm c}$ symmetry which rotates $\Phi$ and $\Phi^c$ is 
broken by the hypercharge gauge interactions of Eqn.~(\ref{dcov-general}) and 
the T term of Eqn.~(\ref{tee-general}).

We present a list of gauge-invariant chiral operators through energy-dimension 
four which conserve CP.  There are twelve such operators given 
by\cite{lon81,aw93}
\begin{subequations}
\label{twelve-chiral}
\begin{eqnarray}
\label{l1p}
{\cal L}_1^\prime 
  & = & \frac{1}{4}\beta_1 v^2 \Big[ \tr ( T V_\mu ) \Big]^2 \;,\\
\label{l1}
{\cal L}_1 & = & \frac{1}{2}\alpha_1 g g^\prime \tr 
                 \Big(B_{\mu\nu} T W^{\mu\nu} \Big) \;,\\
\label{l2}
{\cal L}_2 & = & \frac{i}{2} \alpha_2 g^\prime
	B_{\mu\nu} \tr \Big( T [ V^\mu , V^\nu ] \Big) \;,\\
\label{l3}
{\cal L}_3 & = & i \alpha_3 g \tr \Big( W_{\mu\nu} [V^\mu , V^\nu ] \Big) \;,\\
\label{l4}
{\cal L}_4 & = & \alpha_4 \Big[ \tr ( V_\mu V_\nu ) \Big]^2 \;,\\
\label{l5}
{\cal L}_5 & = & \alpha_5 \Big[ \tr ( V_\mu  V^\mu ) \Big]^2 \;,\\
\label{l6}
{\cal L}_6 & = & \alpha_6 \tr \Big( V_\mu V_\nu \Big) 
	\tr \Big( T V^\mu \Big) \tr \Big( T V^\nu \Big) \;,\\
\label{l7}
{\cal L}_7 & = & \alpha_7 \tr \Big( V_\mu V^\mu \Big) 
	\tr \Big( T V_\nu \Big) \tr \Big( T V^\nu \Big) \;,\\
\label{l8}
{\cal L}_8 & = & \frac{1}{4}\alpha_8 g^2 \Big[ \tr(TW_{\mu\nu})\Big]^2 \;,\\
\label{l9}
{\cal L}_9 & = & \frac{i}{2}\alpha_9 g \tr \Big( T W_{\mu\nu} \Big) 
	\tr \Big( T [V^\mu , V^\nu ] \Big) \;,\\
\label{l10}
{\cal L}_{10} & = & \frac{1}{2} \alpha_{10}
	\Big[ \tr ( T V_\mu ) \tr ( T V_\nu ) \Big]^2 \;,\\
\label{l11}
{\cal L}_{11} & = & \alpha_{11} g\, \epsilon^{\mu\nu\rho\sigma}
\tr \Big( T V_\mu \Big) \tr \Big( V_\nu W_{\rho\sigma} \Big) \;.
\end{eqnarray}
\end{subequations}
The dimension-two operator ${\cal L}_1^\prime$ and the first ten 
dimension-four operators, ${\cal L}_1$ through ${\cal L}_{10}$, conserve both 
C and P, whereas the last operator, ${\cal L}_{11}$, is both C-odd and P-odd.  
We adopt the notation of Ref.~\cite{lon81} and 
Ref.~\cite{aw93}\footnote{Operators ${\cal L}_{1}$ through ${\cal L}_{10}$
were discussed in Ref.~\protect\cite{lon81}, but ${\cal L}_{11}$ was added 
in Ref.~\protect\cite{aw93}}.

In Table~\ref{table-chiral} we indicate the vertices to which each operator 
contributes with an `{\cal X}'.  Additionally we present, with each chiral 
operator, its counterpart in the linear realization.  In particular we may 
associate four of the  chiral operators with energy-dimension-six
operators of Section~\ref{sec-linear}.  Realizing that the ${\cal O}_i$
depend explicitly upon the field $H$, but the ${\cal L}_i$ do not, we may 
write 
\begin{subequations}
\label{relate4}
\begin{eqnarray}
\label{l1p0p1}
{\cal L}_1^\prime & = & - \frac{4 \beta_1}{v^2} {\cal O}_{\Phi,1} \;, \\
\label{l1obw}
{\cal L}_1 & = &  \frac{4 \alpha_1}{v^2} {\cal O}_{BW} \;, \\
\label{l2ob}
{\cal L}_2 & = &  \frac{8 \alpha_2}{v^2} {\cal O}_{B} \;, \\
\label{l3ow}
{\cal L}_3 & = &   \frac{8 \alpha_3}{v^2} {\cal O}_{W} \;.
\end{eqnarray}
\end{subequations}
These operator identities give valid relations among matrix elements for 
processes that do not involve external Higgs particles.
The linear-realization counterparts of the remaining chiral operators appear
at energy-dimension eight, ten and twelve. These higher dimensional operators 
in the second column of Table~\ref{table-linear} are\cite{hag95}
\begin{subequations}
\begin{eqnarray}
{\cal O}^{(8)}_{4} & = & \Big[ ( \dcov_{\mu} \Phi )^\dagger 
( \dcov_{\nu} \Phi ) + 
( \dcov_{\nu} \Phi )^\dagger ( \dcov_{\mu} \Phi ) \Big]^2 \;,\\
{\cal O}^{(8)}_{5} & = & \Big[ ( \dcov_{\mu} \Phi )^\dagger 
( \dcov^{\mu} \Phi ) \Big]^2 \;,\\
{\cal O}^{(10)}_{6} & = & \Big[ ( \dcov_{\mu} \Phi )^\dagger 
( \dcov_{\nu} \Phi ) \Big] \, \Big[ \Phi^\dagger ( \dcov^{\mu} \Phi ) \Big]\,
\Big[ \Phi^\dagger ( \dcov^{\nu} \Phi ) \Big] \;,\\
{\cal O}^{(10)}_{7} & = & \Big[ ( \dcov_{\mu} \Phi )^\dagger 
( \dcov^{\mu} \Phi ) \Big] \, \Big[ \Phi^\dagger ( \dcov_{\nu} \Phi ) \Big]\,
\Big[ \Phi^\dagger ( \dcov^{\nu} \Phi ) \Big] \;,\\
{\cal O}^{(8)}_{8} & = & \Big[ \Phi^\dagger \hat{W}_{\mu\nu} \Phi \Big]^2 \;,\\
{\cal O}^{(8)}_{9} & = & \Big[ \Phi^\dagger \hat{W}_{\mu\nu} \Phi \Big]
\Big[ ( \dcov^{\mu} \Phi )^\dagger 
( \dcov^{\nu} \Phi ) \Big] \;,\\
{\cal O}^{(12)}_{10} & = & \bigg( \Big[ \Phi^\dagger ( \dcov_{\mu} \Phi ) \Big]\,
\Big[ \Phi^\dagger ( \dcov_{\nu} \Phi ) \Big] \bigg)^2 \;.\\
{\cal O}^{(8)}_{11} & = & i \epsilon^{\mu\nu\rho\sigma} 
\Big[ \Phi^\dagger ( \dcov_{\mu} \Phi ) \Big]\,
\Big[ \Phi^\dagger \hat{W}_{\rho\sigma} ( \dcov_{\nu} \Phi ) \Big] + \hc 
\end{eqnarray}
\end{subequations}
\begin{table}[htbp]
\begin{tabular}{|l||c||c|c|c|c|c|c|c|c|c|c|c|}
\makebox[1cm]{}\raisebox{8mm}{${\cal L}_{\rm chiral}$} &
\makebox[.2cm]{}\raisebox{8mm}{${\cal O}^{(n)}_{\rm linear}$} & 
\rotate[l]{\makebox[2cm][c]{WW}} & \rotate[l]{\makebox[2cm][c]{ZZ}} & 
\rotate[l]{\makebox[2cm][c]{AZ}} & \rotate[l]{\makebox[2cm][c]{AA}} & 
\rotate[l]{\makebox[2cm][c]{WWZ}} & \rotate[l]{\makebox[2cm][c]{WWA}} & 
\rotate[l]{\makebox[2cm][c]{WWWW}} & \rotate[l]{\makebox[2cm][c]{WWZZ}} & 
\rotate[l]{\makebox[2cm][c]{WWZA}} & \rotate[l]{\makebox[2cm][c]{WWAA}} & 
\rotate[l]{\makebox[2cm][c]{ZZZZ}} \\ \hline \hline
\raisebox{-0.2cm}{${\cal L}_1^\prime 
   = \frac{\beta_1 v^2}{4} \Big[ \tr ( T V_\mu ) \Big]^2 $} &
\raisebox{-0.2cm}{$-\frac{4 \beta_1}{v^2} {\cal O}_{\Phi,1}$} 
&\makebox[0.5cm]{}& \cm &&&&&&&&\makebox[0.5cm]{}& \\[0.4cm] \hline
\raisebox{-0.2cm}{${\cal L}_1 = \frac{\alpha_1 g g^\prime}{2} B_{\mu\nu} 
\tr \Big( T W^{\mu\nu} \Big)$} &
\raisebox{-0.2cm}{$\frac{4 \alpha_1}{v^2} {\cal O}_{BW}$} & 
& \cm & \cm & \cm & \cm & \cm &&&&& \\[0.4cm] \hline
\raisebox{-0.2cm}{${\cal L}_2 = \frac{i \alpha_2 g^\prime}{2} B_{\mu\nu}
\tr \Big( T [ V^\mu , V^\nu ] \Big) $} &
\raisebox{-0.2cm}{$\frac{8 \alpha_2 }{v^2} {\cal O}_{B}$} & 
&&&& \cm & \cm &&&&& \\[0.4cm] \hline
\raisebox{-0.2cm}{${\cal L}_3 = i \alpha_3 g \,
\tr \Big( W_{\mu\nu} [ V^\mu , V^\nu ] \Big)  $} &
\raisebox{-0.2cm}{$\frac{8 \alpha_3}{v^2} {\cal O}_{W}$} & 
&&&& \cm & \cm & \cm & \cm & \cm && \\[0.4cm] \hline
\raisebox{-0.2cm}{${\cal L}_4 = \alpha_4 \Big[ \tr ( V_\mu V_\nu ) \Big]^2 $} &
\raisebox{-0.2cm}{$\frac{4 \alpha_4}{v^4} {\cal O}^{(8)}_{4}$} & 
&&&&&& \cm & \cm &&& \cm \\[0.4cm] \hline
\raisebox{-0.2cm}{${\cal L}_5 = \alpha_5 \Big[ \tr ( V_\mu  V^\mu ) \Big]^2 $} &
\raisebox{-0.2cm}{$\frac{16 \alpha_5}{v^4} {\cal O}^{(8)}_{5}$} & 
&&&&&& \cm & \cm &&& \cm \\[0.4cm] \hline
\raisebox{-0.2cm}{${\cal L}_6 = \alpha_6 \tr \Big( V_\mu V_\nu \Big) 
\tr \Big( T V^\mu \Big) 
\tr \Big( T V^\nu \Big) $} &
\raisebox{-0.2cm}{$-\frac{64 \alpha_6}{v^6} {\cal O}^{(10)}_{6}$} & 
&&&&&&& \cm &&& \cm \\[0.4cm] \hline
\raisebox{-0.2cm}{${\cal L}_7 = \alpha_7 \tr \Big( V_\mu V^\mu \Big) 
\tr \Big( T V_\nu \Big) 
\tr \Big( T V^\nu \Big) $} &
\raisebox{-0.2cm}{$-\frac{64 \alpha_7}{v^6} {\cal O}^{(10)}_{7}$} & 
&&&&&&& \cm &&& \cm \\[0.4cm] \hline
\raisebox{-0.2cm}{${\cal L}_8 = \frac{\alpha_8 g^2}{4} 
\Big[ \tr ( T W_{\mu\nu} ) \Big]^2 $} &
\raisebox{-0.2cm}{$-\frac{4 \alpha_8 } {v^4} {\cal O}^{(8)}_{8}$} & 
& \cm & \cm & \cm & \cm & \cm & \cm &&&& \\[0.4cm] \hline 
\raisebox{-0.2cm}{${\cal L}_9 = \frac{i \alpha_9 g}{2} 
\tr \Big( T W_{\mu\nu} \Big) 
\tr \Big( T [V^\mu , V^\nu ] \Big) $} &
\raisebox{-0.2cm}{$-\frac{16 \alpha_9 }{v^4} {\cal O}^{(8)}_{9}$} & 
&&&& \cm & \cm & \cm &&&& \\[0.4cm] \hline  
\raisebox{-0.2cm}{${\cal L}_{10} = \frac{\alpha_{10}}{2} \Big[ 
\tr ( T V_\mu ) \tr ( T V_\nu ) \Big]^2 $} &
\raisebox{-0.2cm}{$\frac{128\, \alpha_{10} }{v^8} {\cal O}^{(12)}_{10}$} & 
&&&&&&&&&& \cm \\[0.4cm] \hline
\raisebox{-0.2cm}{${\cal L}_{11} = \alpha_{11}\, g\, \epsilon^{\mu\nu\rho\sigma}
\tr \Big( T V_\mu \Big) \tr \Big( V_\nu W_{\rho\sigma} \Big)  $} &
\raisebox{-0.2cm}{$\frac{8 \alpha_{11}}{v^4} {\cal O}^{(8)}_{11}$} & 
&&&& \cm &&&& \cm && \\[0.4cm]
\end{tabular}
\vspace{.2cm}
\caption{Column one lists op\-er\-a\-tors in the non\-lin\-e\-ar 
re\-pre\-sen\-ta\-tion.  The lin\-e\-ar-re\-pre\-sen\-ta\-tion count\-er\-parts 
appear in the
second column.   For the definitions of the operators ${\cal O}^{(n)}_{i}$
the reader is referred to the text.
An \bf `X' \rm is used to indicate the the contribution of an individual 
operator to a particular vertex.}
\label{table-chiral}
\end{table}
The higher dimensionality of the associated operators in the linear realization
indicates that the observation of effects arising from ${\cal L}_4$ through 
${\cal L}_{11}$ are an indication of a strongly interacting Higgs sector.

Three of the operators, ${\cal L}_{1}^\prime$, ${\cal L}_{1}$ and 
${\cal L}_{8}$, contribute to gauge-boson two-point-functions.  Like 
${\cal O}_{\Phi,1}$, ${\cal L}_{1}^\prime$ contributes only to the $Z$-mass term 
but not to the $W$-mass term and leads to a violation of the custodial symmetry.
Through contributions to the charge form-factors $\baresq(q^2)$, $\barssq(q^2)$
and $\bargzsq(q^2)$ these three operators contribute to 
the process \eeww.\/  None of the operators contributes to the $WW$ 
two-point-function, hence, in contrast to the linear realization, the non-SM
operators do not contribute to the $W$-boson wave-function-renormalization 
factor, and the t-channel neutrino-exchange amplitudes are not modified.

In total six of the operators, ${\cal L}_{1}$, ${\cal L}_{2}$, ${\cal L}_{3}$,
${\cal L}_{8}$, ${\cal L}_{9}$ and ${\cal L}_{11}$,  contribute directly to 
three-gauge-boson vertices, and a total of nine contribute to 
four-gauge-boson vertices.  While operators of the linear representation 
contribute to 
the $WW\gamma\gamma$ vertex but not to a $ZZZZ$ vertex, precisely the opposite 
is realized in Table~\ref{table-chiral}.  And of course there are no Higgs-boson 
interactions in the nonlinear realization.

Finally, an alternative standard notation of Ref.~\cite{gl84andgl85} is related
to our notation by
\begin{subequations}
\label{gasser-leutwyler}
\begin{eqnarray}
\label{gl-L10}
\alp{1} & = & L_{10}
\;, \\
\label{gl-L9R}
\alp{2} & = & -\frac{1}{2}L_{9R}
\;, \\
\label{gl-L9L}
\alp{3} & = & -\frac{1}{2}L_{9L}
\;.
\end{eqnarray}
\end{subequations}
Ref.~\cite{ht90} makes the estimate
\begin{subequations}
\begin{equation}
\label{L10-techni8x8}
\alpha_1 = L_{10} \approx -0.05
\end{equation}
for one family of techniquarks and technileptons with chiral 
SU(8)$\times$SU(8) symmetry, and
\begin{equation}
\label{L10-techni-singlet}
\alpha_1 = L_{10} \approx -0.005
\end{equation}
in the minimal model with one color-singlet technidoublet.  Taking input from 
low-energy QCD\cite{gl84andgl85,hol91},
\begin{equation}
\label{L9=-L10}
L_9 \approx - L_{10}\;,
\makebox[.5cm]{} \alpha_2 \approx \alpha_3 \approx 2 \alpha_1   \;.
\end{equation}
\end{subequations}
Considering the contributions from $N$ flavor doublets of heavy fermions $U$ 
and $D$ \cite{hol91,apls88,tc} Ref.~\cite{aw93} 
estimates 
\begin{subequations}
\label{chiral-quark-est}
\begin{eqnarray}
\label{aw-est-beta1}
\beta_1 & = & \frac{N}{24\pi^2} \frac{(\Delta m)^2}{v^2} \;, \\
\label{aw-est-alpha1}
\alpha_1 & = & - \frac{N}{96\pi^2} \approx -N \times 10^{-3}\;, \\
\label{aw-est-alpha2}
\alpha_2 & = &  - \frac{N}{96\pi^2} \approx -N \times 10^{-3}\;, \\
\label{aw-est-alpha3}
\alpha_3 & = & - \frac{N}{96\pi^2}\bigg\{1-\frac{2}{5}\delta^2\bigg\}
             \approx -N \times 10^{-3} \;, \\
\label{aw-est-alpha8}
\alpha_8 & = & - \frac{N}{96\pi^2}\frac{16}{5}\delta^2\;, \\
\label{aw-est-alpha9}
\alpha_9 & = & - \frac{N}{96\pi^2}\frac{14}{5}\delta^2\;, \\
\label{aw-est-alpha11}
\alpha_{11} & = & \frac{N}{96\pi^2}\delta\;,
\end{eqnarray}
\end{subequations}
where $\Delta m = m_U - m_D$, $\delta = (m_U - m_D)/(m_U + m_D)$, and it has 
been assumed that the mass splitting is small compared to the masses $m_U$ and 
$m_D$.  Notice that $\alpha_1$, $\alpha_2$ and $\alpha_3$ are approximately
degenerate.  Also, $\beta_1$ is suppressed relative to $\alpha_1$ by 
$(\Delta m)^2/v^2$ while $\alpha_8$ and $\alpha_9$ are suppressed by $\delta^2$.
It is noteworthy that, while $\alpha_{11}$ is also suppressed, the suppression 
factor is only one power of $\delta$.  The above estimates will serve as useful
benchmarks throughout the remainder of the paper.

 
\section{Processes with four external fermions}\label{sec-eeff}
\subsection{The linear realization of the symmetry-breaking sector}
\label{sec-eeff-linear}

In the linear realization four operators, ${\cal O}_{DW}$, ${\cal O}_{DB}$, 
${\cal O}_{BW}$ and ${\cal O}_{\Phi,1}$, have special 
significance\cite{gw91,hisz92,hisz93,hms95}  due to their contributions to 
low-energy 
processes involving four external light fermions.  In short, they are the only 
operators from Eqns.~(\ref{twelve-operators}) which are well constrained by the 
present data.  This subset contributes to electroweak precision observables via 
their contributions to the transverse components of the gauge-boson propagators. 
If the one-particle-irreducible two-point-function is separated into SM and 
new-physics contributions according to $ \overline{\Pi} = \overline{\Pi}_{\rm SM}
 + \Delta\overline{\Pi} $, then, in the notation of Ref.~\cite{hisz92,hisz93}, 
we find
\begin{subequations}
\label{two_point_functions}
\begin{eqnarray} \label{piqq}
\Delta\overline{\Pi}^{QQ}_T(q^2) &=&  2 \frac{q^2}{\Lambda^2} 
\biggl[ (f_{DW}+ f_{DB}) q^2  - f_{BW} \frac{v^2}{4} \biggr]\; ,
\\ \label{pi3q}
\Delta\overline{\Pi}^{3Q}_T(q^2) &=& 2 \frac{q^2}{\Lambda^2} \biggl[ f_{DW} q^2
 - f_{BW} \frac{v^2}{8} \biggr]  \; ,
\\ \label{pi33}
\Delta\overline{\Pi}^{33}_T(q^2) &=& 2 \frac{q^2}{\Lambda^2} f_{DW} q^2
 -  \frac{v^2}{\Lambda^2} \biggl[ f_{\Phi,1} + f_{\Phi,4} \biggr] 
  \frac{v^2}{8} \; , 
\\ \label{pi11}
\Delta\overline{\Pi}^{11}_T(q^2) &=& 2 \frac{q^2}{\Lambda^2} f_{DW} q^2 
  -  \frac{v^2}{\Lambda^2} f_{\Phi,4} \frac{v^2}{8} \; .
\end{eqnarray}
\end{subequations}
The two-point functions may also be expressed in a basis which refers to 
physical gauge bosons by

\begin{subequations}
\label{two_point_functions_physical}
\begin{eqnarray} 
\label{pigg}
\overline{\Pi}^{\gamma\gamma}_T(q^2) & = & \hatesq  \overline{\Pi}^{QQ}_T(q^2) 
\;, \\ 
\label{pigz}
 \overline{\Pi}^{\gamma Z}_T(q^2) & = & \hate\hatgz \bigg[ 
\overline{\Pi}^{3Q}_T(q^2) - \hatssq  \overline{\Pi}^{QQ}_T(q^2) \bigg]\;,
\\ 
\label{pizz}
 \overline{\Pi}^{ZZ}_T(q^2) & = & \hatgzsq \bigg[  \overline{\Pi}^{33}_T(q^2) 
-2\hatssq  \overline{\Pi}^{3Q}_T(q^2) + \hat{s}^4  \overline{\Pi}^{QQ}_T(q^2) 
\bigg]\;,
\\ 
\label{piww}
 \overline{\Pi}^{WW}_T(q^2) & = & \hatgsq  \overline{\Pi}^{11}_T(q^2) \;.
\end{eqnarray}
\end{subequations}
Either set of two-point functions may be employed, as convenience dictates.
From Eqn.~(\ref{two_point_functions}) follow the $S$, $T$ and $U$ 
parameters of Ref.~\cite{pt90andpt92} or some equivalent triplet of 
parameters\cite{others}.  In general we allow for an 
anomalous contribution to $\alpha_{\rm QED}(m_Z^2)$\cite{kni94}.
Defining $S$, $T$ and $U$ according to Ref.~\cite{hhkm94}, 
\begin{subequations}
\label{relationship}
\begin{eqnarray}\label{deltas}
\Delta S & \equiv & 16\pi\,{\cal R}\!e
\bigg[\Delta\overline{\Pi}^{3Q}_{T,\gamma}(m^2_Z) - 
\Delta\overline{\Pi}^{33}_{T,Z}(0)\bigg]
= - 4\pi \frac{v^2}{\Lambda^2}f_{BW}
\;,
\\ \label{deltat}
\Delta T & \equiv & \frac{4\sqrt{2} G_F}{\alpha} \,{\cal R}\!e
\bigg[\Delta\overline{\Pi}^{33}_{T}(0) - \Delta\overline{\Pi}^{11}_{T}(0)\bigg]
= - \frac{1}{2 \alpha} \frac{v^2}{\Lambda^2}f_{\Phi,1}
\;,
\\ \label{deltau}
\Delta U & \equiv & \makebox[0.12cm]{}
 16\pi\,{\cal R}\!e
\bigg[\Delta\overline{\Pi}^{33}_{T,Z}(0) - \Delta\overline{\Pi}^{11}_{T,W}(0)\bigg]
\makebox[0.13cm]{}
= 32\pi \frac{m_Z^2-m_W^2}{\Lambda^2}f_{DW}
\;,
\\ \label{deltaalpha}
\Delta\frac{1}{\alpha} & \equiv & \makebox[0.12cm]{}
 4\pi\,{\cal R}\!e
\bigg[\Delta\overline{\Pi}^{QQ}_{T,\gamma}(m^2_Z) - \Delta\overline{\Pi}^{QQ}_{T,\gamma}(0)\bigg]
\makebox[0.13cm]{} =
8\pi \frac{m_Z^2}{\Lambda^2} \Big(f_{DW} + f_{DB}\Big)
\;,
\end{eqnarray}
\end{subequations}
where $S=S_{\rm SM} + \Delta S$, $T=T_{\rm SM} + \Delta T$, $U=U_{\rm SM} + 
\Delta U$, and
\begin{eqnarray}
\overline{\Pi}^{AB}_{T,V}(q^2) = {\overline{\Pi}^{AB}_T(q^2) - \overline{\Pi}^{AB}_T(m^2_V)\over q^2 -
m^2_V}\;.
\end{eqnarray}
Because the contributions of $f_{\Phi,1}$ and $f_{\Phi,4}$ to the two-point 
functions of Eqn.~(\ref{two_point_functions}) are independent of $q^2$, they 
may contribute only to $T$.  The $f_{\Phi,4}$ contributions exactly cancel
as expected.  The charge form-factors of Ref.~\cite{hhkm94} follow directly;
\begin{subequations}
\label{gzb20_sb2mz2_gwb20}
\begin{eqnarray} 
\label{gzb20}
\frac{1}{ \overline{g}_Z^2(0) }& = & \frac{ 1 + \overline{\delta_G} 
- \alpha T}{ 4 \sqrt{2}
G_F m_Z^2}\;, \\
\label{sb2mz2}
\overline{s}^2(m_Z^2) & = & \frac{1}{2} - \sqrt{ \frac{1}{4} -  
\overline{e}^2(m_Z^2)\left( \frac{1}{ \overline{g}_Z^2(0) } 
+ \frac{S}{16\pi}\right)}\;, \\
\label{gwb20}
\frac{1}{ \overline{g}_W^2(0) }& = & \frac{\overline{s}^2(m_Z^2)}{
\overline{e}^2(m_Z^2)} - \frac{1}{16\pi} \left(S+U\right)\;,
\end{eqnarray}
\end{subequations}
where SM vertex and box corrections to the muon lifetime are incorporated in 
$\overline{\delta_G} \approx 0.0055$.  Additionally, the nontrivial 
$q^2$-dependence of the two-point-functions leads to a nonstandard running of 
the charge form-factors;
\begin{subequations}
\label{running}
\begin{eqnarray}
\label{run_alpha}\Delta\Bigg[
\frac{1}{\overline{e}^2(q^2)} - \frac{1}{4\pi\alpha} \Bigg]&=& 
2 \frac{q^2}{\Lambda^2} \Big(f_{DW} + f_{DB}\Big)\;,\\
\label{run_sb2}\Delta\Bigg[
\frac{\overline{s}^2(q^2)}{\overline{e}^2(q^2)}
- \frac{\overline{s}^2(m_Z^2)}{\overline{e}^2(m_Z^2)}\Bigg] & = & 
2 \frac{q^2 - m_Z^2}{\Lambda^2} f_{DW}\;,\\
\label{run_gzb2}\Delta\Bigg[
\frac{1}{ \overline{g}_Z^2(q^2) } - \frac{1}{ \overline{g}_Z^2(0) }\Bigg] & = &
2 \frac{q^2}{\Lambda^2}\Big( \hat{c}^4 f_{DW} + \hat{s}^4 f_{DB} \Big)\;,\\
\label{run_gwb2}\Delta\Bigg[
\frac{1}{ \overline{g}_W^2(q^2) } - \frac{1}{ \overline{g}_W^2(0) }\Bigg] & = &
2 \frac{q^2}{\Lambda^2}f_{DW}\;.
\end{eqnarray}
\end{subequations}
The combination of Eqn.~(\ref{gzb20_sb2mz2_gwb20}) with Eqn.~(\ref{running})
leads to the convenient expressions
\begin{subequations}
\label{corrections}
\begin{eqnarray}\label{crctn_alpha}
\Delta \overline{\alpha}(q^2) & = & 
      - 8 \pi \hat{\alpha}^2 
	\frac{q^2}{\Lambda^2}\Big( f_{DW} + f_{DB} \Big)\;,
\\ \label{crctn_gzbar2}
\Delta \overline{g}_Z^2(q^2) & = & 
  	- 2 \hat{g}_Z^4 \frac{q^2}{\Lambda^2} \Big(\hat{c}^4 f_{DW} 
	+ \hat{s}^4 f_{DB} \Big) 
 	- \frac{1}{2} \hat{g}_Z^2 \frac{v^2}{\Lambda^2} f_{\Phi,1} \;,
\\ \nonumber
\Delta \overline{s}^2(q^2) & = & 
	 \frac{-\hat{s}^2\hat{c}^2}{\hat{c}^2-\hat{s}^2}\Bigg[ 
	8\pi\hat{\alpha}\frac{m_Z^2}{\Lambda^2}\Big( f_{DW} + f_{DB} \Big)
	+ \frac{m_Z^2}{\Lambda^2} f_{BW} 
	- \frac{1}{2}\frac{v^2}{\Lambda^2}f_{\Phi,1}
	\Bigg]
\\
\label{crctn_sbar2}
   	&& \makebox[5.5cm]{} + 8\pi\hat{\alpha}\frac{q^2-m_Z^2}{\Lambda^2} 
   	  \Big( \hat{c}^2 f_{DW} - \hat{s}^2 f_{DB}\Big)\;,
\\ \label{crctn_gwbar2}
\Delta \overline{g}_W^2(q^2) & = & 
	-8\pi\hat{\alpha}\hat{g}^2 \frac{m_Z^2}{\Lambda^2}f_{DB}
	- \hat{g}^2 \frac{ \Delta \overline {s}^2(m_Z^2)}{\hat{s}^2} 
	- \frac{1}{4} \hat{g}^4 \frac{v^2}{\Lambda^2}f_{BW} 
	- 2\hat{g}^4 \frac{q^2}{\Lambda^2}f_{DW} \;.
\end{eqnarray}
\end{subequations}
The `hatted' couplings are the $\overline{\rm MS}$ couplings, and hence they 
satisfy the tree-level relationships $\hat{e} \equiv \hat{g}\hat{s} \equiv 
\hat{g}_Z\hat{s}\hat{c}$ and $\hat{e}^2 \equiv 4\pi\hat{\alpha}$.  For 
numerical results concerning $Z$-pole observables we adopt the renormalization 
conditions of Ref.~\cite{kni94} and use $\baralpha(\mzsq)_{\rm SM} =128.72$ 
and $\barssq(\mzsq)_{\rm SM} = 0.2312$.

We perform a $\chi^2$ analysis to constrain the corrections of 
Eqn.~(\ref{corrections}).  We base our analysis on the results of the recent 
global analysis of Ref.~\cite{hhkm94,hhm96}.  
In Ref.~\cite{hhm96} the `barred' charges are fit to the data with the following 
results.  For measurements on the $Z$-pole,
\begin{subequations}
\label{z-pole-data}
\begin{eqnarray}
\label{z-pole-gzb2-data}
\bargzsq(\mzsq) & = & 0.55557 - 0.00042
\frac{\alpha_s + 1.54 \overline{\delta}_b(\mzsq) - 0.1065}{0.0038} \pm 0.00061 
\;, \\
\label{z-pole-sb2-data}
\barssq(\mzsq) & = & 0.23065 + 0.00003
\frac{\alpha_s + 1.54 \overline{\delta}_b(\mzsq) - 0.1065}{0.0038} \pm 0.00024
\;, \\
\label{z-pole-corr-data}
\rho_{\rm corr} & = & 0.24 \;.
\end{eqnarray}
\end{subequations}
The parameter $\overline{\delta}_b(\mzsq)$, which is a function of $m_t$, 
accounts for corrections to the $Zb\overline{b}$ vertex, for which adopt the SM 
values.  Combining the $W$-boson mass measurement 
($m_W = 80.356\pm 0.125$GeV) 
with the input parameter $G_F$, they find 
\begin{eqnarray}
\label{w-data}
\bargwsq(0) = 0.4237 \pm 0.0013\;.
\end{eqnarray}
And finally, from the low-energy data,
\begin{subequations}
\label{le-data}
\begin{eqnarray}
\label{le-gzb2-data}
\bargzsq(0) & = & 0.5441 \pm 0.0029
\;, \\
\label{le-sb2-data}
\barssq(0) & = & 0.2362 \pm 0.0044
\;, \\
\label{le-corr-data}
\rho_{\rm corr} & = & 0.70 \;.
\end{eqnarray}
\end{subequations}
For $\alpha_s = 0.118$ we obtain the following constraints on $\fdw$, $\fdb$,
$\fbw$ and $\fpone$: 
\begin{eqnarray}
\label{fit_now}
\left.
\begin{array}{lll}
f_{DW} & = & -0.32 + 0.0088 \,x_H - 0.55 \,x_t \pm 0.44 \\[1mm]
f_{DB} & = & -14 \pm 10 \\[1mm]
f_{BW} & = & 3.7  + 0.085 \,x_H \pm 2.4 \\[1mm]
f_{\Phi,1} & = & 0.30 - 0.028 \,x_H + 0.32 \,x_t  \pm 0.16 
\end{array}
\right.
\makebox[2mm]{}
\left(\makebox[-0.3cm]{}
\begin{array}{lddd}
\dec 1. & \dec $-$0.191 & \dec    0.055 & \dec $-$0.237  \\[1mm]
        & \dec    1.    & \dec $-$0.988 & \dec $-$0.884  \\[1mm]
        &               & \dec    1.    & \dec    0.943  \\[1mm]
        &               &               & \dec    1.
\end{array} \right) 
\end{eqnarray}
where 
\begin{equation}
x_t = \frac{m_t-175{\rm GeV}}{100{\rm GeV}}\;,  \makebox[1cm]{}
x_H = \ln \frac{m_H}{100{\rm GeV}} \;,
\end{equation}
and $\Lambda = 1$TeV.
The errors are at the one-sigma level.  The parameterization of the central 
values is good to a few percent of the one-sigma errors 
in the range $150{\rm GeV}<m_t<190{\rm GeV}$ and 
$60{\rm GeV}<m_H<800{\rm GeV}$; the dependencies upon $m_H$ and $m_t$ arise
from SM contributions only.  

We note the very strong correlations among three of the parameters.  This 
suggests that the data constrains one combination of the parameters particularly 
well.  This should not be ignored since this most stringent constraint sets the 
present sensitivity limit for physics beyond the SM as parameterized by the 
effective Lagrangian of Eqn.~(\ref{fulllagrangian}).  We diagonalize the 
covariance matrix and repeat the $\chi^2$ analysis in the basis of eigenvectors 
to find this particular combination and its associated error with the following 
result:
\begin{equation}
\label{best-combo}
\fpone - 0.18 \fbw - 0.029 \fdb + 0.016 \fdw = 0.023 \pm 0.017 \;.
\end{equation}
The implication of this measurement is that, barring accidental cancellations 
among the various parameters, the constraint on $\fpone$ is actually much
more severe than one would expect from Eqn.~(\ref{fit_now}).

This result may be explained by the dominance of the data from the $Z$-pole 
measurements.  Comparing the errors associated with the charge form-factors of
Eqn.~(\ref{z-pole-data}), Eqn.~(\ref{w-data}) and Eqn.~(\ref{le-data}), it is 
clear that the measurements of $\barssq(\mzsq)$ and $\bargzsq(\mzsq)$ are much
more precise than the remaining measurements.  Considering only these two 
measurements and including their associated correlation it is possible to 
predict which combination of parameters is best constrained, and that prediction
is approximately Eqn.~(\ref{best-combo}).  Nevertheless, the low-energy 
neutral-current and the charged-current/$W$-boson-mass data play an important 
role in the fit.

\subsection{The nonlinear realization of the symmetry-breaking sector}
\label{sec-eeff-nonlinear}

We may repeat the entire analysis for the chiral Lagrangian of 
Eqn.~(\ref{fulllagrangiannl}).  The corrections to the two-point-functions are
\begin{subequations}
\label{two_point_functions_nl}
\begin{eqnarray} 
\label{piqq_nl}
\Delta\overline{\Pi}^{QQ}_T(q^2) &=& -q^2\Big( 2\alpha_1 + \alpha_8 \Big) \;, \\ 
\label{pi3q_nl}
\Delta\overline{\Pi}^{3Q}_T(q^2) &=& -q^2\Big( \alpha_1 + \alpha_8 \Big) \;, \\ 
\label{pi33_nl}
\Delta\overline{\Pi}^{33}_T(q^2) &=& \frac{1}{2}\beta_1 v^2 - q^2\alpha_8 \;, \\ 
\label{pi11_nl}
\Delta\overline{\Pi}^{11}_T(q^2) &=& 0 \;.
\end{eqnarray}
\end{subequations}
A comparison of Eqns.~(\ref{two_point_functions_nl}) with 
Eqns.~(\ref{two_point_functions}) reveals two important differences.  The chiral
Lagrangian leads, at the current level of calculation, to at most linear 
dependence of the two-point functions upon $q^2$.  Also 
$\Delta\overline{\Pi}^{11}_T(q^2)$ vanishes in Eqn.~(\ref{pi11_nl}).  In 
analogy with Eqns.~(\ref{relationship}),
\begin{subequations}
\label{relationship_nl}
\begin{eqnarray}
\label{deltas_nl}
\Delta S & = & -16\pi\alpha_1 \;, \\ 
\label{deltat_nl}
\Delta T & = & \frac{2}{\alpha}\beta_1\;, \\ 
\label{deltau_nl}
\Delta U & = & -16\pi\alpha_8 \;, \\ 
\label{deltaalpha_nl}
\Delta\frac{1}{\alpha} &  = & 0 \;,
\end{eqnarray}
\end{subequations}
which agrees  with Ref.~\cite{aw93}.  The contributions to the charge 
form-factors 
may be calculated via Eqn.~(\ref{gzb20_sb2mz2_gwb20}), but there is no 
additional contribution to the running of the charge form-factors in 
Eqn.~(\ref{running}).  Here the analysis with the operators of the chiral 
Lagrangian through energy dimension four is equivalent to the standard $S$, $T$, 
$U$ analysis.  In short, the results are
\begin{subequations}
\label{corrections_nl}
\begin{eqnarray}
\label{crctn_alpha_nl}
\Delta \overline{\alpha}(q^2) & = & 0 \;, \\
\label{crctn_gzbar2_nl}
\Delta \overline{g}_Z^2(q^2) & = & 2 \hatgzsq \beta_1 \;, \\
\label{crctn_sbar2_nl}
\Delta \overline{s}^2(q^2) & = & -\frac{\hatcsq\hatssq}{\hatcsq-\hatssq}
\Big( 2\beta_1 + \hatgzsq \alpha_1 \Big) \;, \\
\label{crctn_gwbar2_nl}
\Delta \overline{g}_W^2(q^2) & = & -\hatgsq 
\frac{\Delta \overline{s}^2(m_Z^2)}{\hatssq} 
- \hat{g}^4 \Big( \alpha_1 + \alpha_8 \Big)  \;.
\end{eqnarray}
\end{subequations}
A fit to the data as summarized by Eqn.~(\ref{z-pole-data}), 
Eqn.~(\ref{w-data}) and Eqn.~(\ref{le-data}) produces the central values, the 
one-sigma errors and the correlation matrix which follow:
\begin{eqnarray}
\label{fit_now_nl}
\left.
\begin{array}{lll}
\alpha_1 & = & (4.3 -4.8 \,x_t \pm 2.6)\times 10^{-3} \\[1mm]
\beta_1  & = & (0.45 -3.5 \,x_t \pm 0.57)\times 10^{-3} \\[1mm]
\alpha_8 & = & (-0.71 + 9.1 \,x_t \pm 7.6 )\times 10^{-3}
\end{array}
\right.
\makebox[2mm]{}
\left(\makebox[-0.3cm]{}
\begin{array}{lddd}
\dec 1.   & \dec $-$0.87 & \dec $-$0.12 \\[1mm]
          & \dec    1.   & \dec    0.22 \\[1mm]
          &              & \dec    1.    
\end{array} \right) 
\end{eqnarray}
For our reference values of $S$, $T$ and $U$ we use the SM values at 
$m_H = 1$TeV.  Notice that the data favors a positive value of $\alpha_1$
while the estimates of Eqns.~(\ref{L10-techni8x8}), (\ref{L10-techni-singlet})
and (\ref{aw-est-alpha1}) all predict a negative value.

The two fits, (\ref{fit_now}) and (\ref{fit_now_nl}), are not equivalent.  The 
nontrivial running of the charge form-factors introduced by the 
energy-dimension-six operators ${\cal O}_{DW}$ and ${\cal O}_{DB}$ is a 
leading-order effect; similar effects will also be induced by dimension-six
operators of the chiral Lagrangian\cite{blm94andbgklm94} which are neglected 
in the present approximation.  On the other 
hand, the contribution of $\alpha_8$ is equivalent to a dimension-eight effect 
in the linear realization, hence the contribution of its counterpart in the 
linear realization is expected to be suppressed.  A partial comparison may be 
made only in the limit where  $f_{DW}=f_{DB}=0$ and $\alpha_8=0$, which 
corresponds to a fit in $\Delta S$ and $\Delta T$ with $\Delta U=0$\cite{hms95}.

\subsection{Expectations for improved measurements}
\label{sec-eeff-future}

The study of four-fermion processes at higher energies will do little to further 
constrain the parameters of the chiral Lagrangian via contributions to the 
electroweak charge form-factors unless the precision of those high-energy 
experiments is competitive with the precision of LEP/SLC.  The situation is 
markedly different when the symmetry breaking is linearly realized, and the 
anomalous running of the charge form-factors leads to enhanced sensitivity at 
higher center of mass energies.  This enhanced sensitivity in turn implies 
improved constraints upon the contributing coefficients.  At LEP~II, with 
$\sqrt{s} =175$GeV and $\int {\cal L}dt = 500$pb$^{-1}$, the constraints may 
improve as
\begin{eqnarray}
\label{fit_lep2}
\left.
\begin{array}{lll}
f_{DW} & = & -0.07 + 0.032 \,x_H - 0.67 \,x_t \pm 0.22 \\[1mm]
f_{DB} & = & -0.3 + 0.13 \,x_H + 0.83 \,x_t  \pm 1.9 \\[1mm]
f_{BW} & = & 0.19  + 0.050 \,x_H \pm 0.46 \\[1mm]
f_{\Phi,1} & = & 0.052 - 0.032 \,x_H + 0.34 \,x_t  \pm 0.042 
\end{array}
\right.
\makebox[-1mm]{}
\left(\makebox[-0.3cm]{}
\begin{array}{lddd}
\dec 1.   & \dec $-$0.490 & \dec    0.211 & \dec $-$0.182  \\[1mm]
          & \dec    1.    & \dec $-$0.896 & \dec $-$0.484  \\[1mm]
          &               & \dec    1.    & \dec    0.791  \\[1mm]
          &               &               & \dec    1.
\end{array} \right) 
\end{eqnarray}
We make the assumption that the measurement of the $W$-boson mass  will 
improve to $\Delta m_W = 45$MeV\cite{lepii-mw-wg}.  
The corresponding improvement for the 
parameters of the chiral Lagrangian is much more modest.  Only the $W$-boson 
mass measurement plays a role, reducing the error on $\alpha_8$ to 
$\delta\alpha_8 = \pm 3.3 \times 10^{-3}$ while increasing the correlation 
between $\alpha_8$ and $\beta_1$ to $0.50$.

At a future linear collider with $\sqrt{s}=500$GeV and $\int {\cal L}dt = 
50$fb$^{-1}$ we may expect
\begin{eqnarray}
\label{fit_nlc}
\makebox[-1.5mm]{}
\left.
\begin{array}{lll}
f_{DW} & = & -0.010 + 0.0089 \,x_H - 0.13 \,x_t \pm 0.055 \\[1mm]
f_{DB} & = & 0.00 - 0.0070 \,x_H \pm 0.21 \\[1mm]
f_{BW} & = & 0.06  + 0.097 \,x_H \pm 0.17 \\[1mm]
f_{\Phi,1} & = & 0.037 - 0.028 \,x_H + 0.34 \,x_t  \pm 0.025 
\end{array}
\right.
\makebox[-1mm]{}
\left(\makebox[-0.3cm]{}
\begin{array}{lddd}
\dec 1.   & \dec 0.295 & \dec $-$0.242 & \dec $-$0.131  \\[1mm]
          & \dec 1.    & \dec $-$0.340 & \dec $-$0.140  \\[1mm]
          &            & \dec    1.    & \dec    0.904  \\[1mm]
          &            &               & \dec    1.
\end{array} \right) 
\end{eqnarray}
In this analysis we have also assumed that the error on the $W$-boson mass
will be reduced to $\Delta m_W = \pm 20$MeV by the TeV33 upgrade of the 
Fermilab Tevatron\cite{tev2000}.  For the parameters of the chiral Lagrangian,
the fit of Eqn.~(\ref{fit_now_nl}) is modified by a reduction of the error on 
$\alpha_8$ to $\delta\alpha_8 = \pm 2.3\times 10^{-3}$ while the correlation 
between $\alpha_8$ and $\beta_1$ is increased to $0.72$.  In this case the 
improvement is from the precise measurement of the $W$-boson mass.

In Ref.~\cite{rv95andrv96andbrtv96} a scheme has been proposed 
which accounts 
for $Z$-pole measurements at LEP~I when discussing new data at LEP~II or a 
higher-energy linear $e^+e^-$ collider.  Their ``$Z$-peak subtracted'' scheme 
reduces the number of parameters required for these future experiments by 
using LEP~I measurements as input parameters for the calculation of observables 
at higher energies; effectively they concentrate on $f_{DW}$ and $f_{DB}$, the 
operators which introduce a nonstandard running of the charge form-factors.
The obvious advantage of their approach is a smaller parameter space which 
focuses on those parameters whose constraints should improve the most.  However,
with an exact calculation we obtain more stringent bounds, and we are able to 
take full advantage of the correlations among all four parameters; these 
correlations change dramatically at different scales.  
Because the details concerning our analyses are quite different, our 
results and theirs are not easily compared.  However, we find rough agreement 
between their results and ours.


\section{The process \eeww}\label{sec-corrections}

Next we calculate the contributions of the effective Lagrangians of 
Section~\ref{sec-linear} and Section~\ref{sec-nonlinear} to the form factors
of Eqn.~(\ref{general-form}) and Eqn.~(\ref{big-ff}).  Eventually it will be
necessary to include both the complete SM radiative corrections and the 
effective-Lagrangian contributions in a combined analysis, but the scenario of
immediate interest is where the nonstandard contributions are relatively large 
compared to the higher-order SM effects.  The SM corrections have been 
considered by many authors\cite{bdsbbk88andfjz89,sm-corr,aow94,lepii-ww-wg}.

When neglecting the SM loop-level corrections, Eqn.~(\ref{big-ff}) may be 
simplified 
considerably.  Because there are no corrections to fermionic vertices, the 
$\Gamma_1^{\,e}$ and $\overline{\Gamma}_2^{\,e}$ terms vanish while the 
$\Gamma^{e\nu}$ term becomes equivalent to the self-energy correction for an 
external $W$ boson.  In the effective Lagrangian the equivalent of a box 
correction is a contact term; there is no such contribution due to the exclusion 
of fermionic fields in the construction of effective operators.  
With these simplifications we may rewrite Eqn.~(\ref{big-ff}) as
\begin{eqnarray}
\nonumber
F^{\rm IB}_{i,\tau}(s,t) & = & \frac{Q}{s}\Bigg\{
	\baresq(s)_{\rm SM} f_i^{\gamma\,(0)} 
	+ \Delta \baresq(s) f_i^{\gamma\,(0)} 
	+ \hatesq f_i^{\gamma\,(1)}(s) \Bigg\}
\\ \nonumber && \makebox[-1.5cm]{} 
	+ \frac{1}{s-m_Z^2 + is\frac{\Gamma_Z}{m_Z}}
	\Bigg\{ \big(I_3-\hatssq Q\big)\hatcsq 
	\bigg( \bargzsq(s)_{\rm SM} f_i^{Z\,(0)}
	+ \Delta \bargzsq(s) f_i^{Z\,(0)}
	+ \hatgzsq f_i^{Z\,(1)}(s) \bigg)
\\ \nonumber && \makebox[5cm]{} 
	- \hatgzsq \bigg( Q \hatcsq f_i^{Z\,(0)} + \big(I_3-\hatssq Q\big) 
	f_i^{\gamma\,(0)} \bigg) \Delta \overline{s}^2(s) \Bigg\}
\\ && \makebox[-1.5cm]{} 
	+ \frac{I_3}{2t} \bigg( \bargwsq(m_W^2)_{\rm SM} 
	+ \Delta \bargwsq(m_W^2) \bigg) 
	f_i^{t\,(0)} \;.
\label{big-simple}
\end{eqnarray}
Notice that, through the $Z$--$\gamma$ mixing term, the $Z$-boson has 
acquired a coupling proportional to the charge of the electron, as discussed 
in Ref.~\cite{hol91}.  Also note that Eqn.~(\ref{big-simple}) remains valid
when the SM corrections to the gauge-boson propagators are included.

\subsection{The linear realization of the symmetry-breaking sector}
\label{sec-eeww-linear}

The corrections to the charge form factors, $\Delta \baresq(q^2)$, 
$\Delta \barssq(q^2)$, $\Delta \bargzsq(q^2)$ and $\Delta \bargwsq(q^2)$, may be 
found in Eqns.~(\ref{corrections}). The tree-level form factors $f_i^{X\,(0)}$ 
may be found in Table~\ref{table-smff}.  Hence, once we calculate the various 
$f_i^{X\,(1)}$ terms, the form factors of Eqn.~(\ref{big-simple}) are 
completely determined.  For the effective Lagrangian of 
Eqn.~(\ref{fulllagrangian}) we find
\begin{subequations}
\label{corr-dir}
\begin{eqnarray}
\label{f1g-dir}
f_1^{\gamma\,(1)}(s) & = & -\hatgsq \frac{s}{\Lamsq}\fdw 
	+ \frac{3}{4} \hatgsq \frac{s}{\Lamsq}\fwww\;,\\
\label{f2g-dir}
f_2^{\gamma\,(1)}(s) & = & -6\hatgsq\frac{\mwsq}{\Lamsq}\fdw 
	+ \frac{3}{2} \hatgsq \frac{\mwsq}{\Lamsq}\fwww\;,\\
\label{f3g-dir}
f_3^{\gamma\,(1)}(s) & = & 2\hatgsq\frac{2s-3\mwsq}{\Lamsq}\fdw
	+ \frac{3}{2}\hatgsq\frac{\mwsq}{\Lamsq}\fwww
	+ \frac{1}{2}\frac{\mwsq}{\Lamsq}\Big(\fw - 2\fbw + \fb\Big)\;,
\end{eqnarray}
for the $WW\gamma$ vertex, and
\begin{eqnarray}
\label{f1z-dir}
f_1^{Z\,(1)}(s) & = & f_1^{\gamma\,(1)}(s) 
	+ \frac{1}{2} \frac{\mzsq}{\Lamsq}\fw\;,\\
\label{f2z-dir}
f_2^{Z\,(1)}(s) & = & f_2^{\gamma\,(1)}(s)\;,\\
\label{f3z-dir}
f_3^{Z\,(1)}(s) & = & f_3^{\gamma\,(1)}(s) 
	+ \frac{1}{2}\frac{\mzsq}{\Lamsq}\Big(\fw + 2\fbw - \fb\Big)\;,
\end{eqnarray}
\end{subequations}
for the $WWZ$ vertex.  Only nonzero results are reported.  Notice that 
$f_1^{Z\,(1)}(s)$ differs from $f_1^{\gamma\,(1)}(s)$ by a term proportional 
to $\fw$; $f_2^{Z\,(1)}(s)$ and $f_2^{\gamma\,(1)}(s)$ are the same; 
the $\fw$, $\fbw$ and $\fb$ terms of $f_3^{\gamma\,(1)}(s)$ and 
$f_3^{Z\,(1)}(s)$ differ.  Recall that the $W$-boson self-energy contributions 
for the external $W$-bosons are included in the $f_i^{X\,(1)}$ form factors.  

Because the effective Lagrangian of Eqn.~(\ref{fulllagrangian}) is invariant 
under U(1)$_{\rm em}$, the $g_1^\gamma(s)$ form-factor is required to assume 
its canonical value, $g_1^\gamma(0)=1$, for on-shell photons.  For readers who 
are unfamiliar with this standard notation\cite{hpzh87}, it will be reviewed
later in this section.  We may obtain the $g_1^\gamma(s)$ form-factor via
\begin{equation}
\label{g1g-general}
g_1^{\gamma}(s) = f_1^{\gamma}(s) - \frac{s}{2\mwsq}f_2^{\gamma}(s)\;.
\end{equation}
However, care must be taken to account for direct corrections to the three-point
vertex as well as self-energy corrections for the particles attached to each leg
of the vertex.  Motivated by the form of the first line of 
Eqn.~(\ref{big-simple}) we define
\begin{equation}
\label{fi-gamma-eff}
f_i^{\gamma\,(\rm eff)}(s) = 
   \bigg[ 1 + \frac{\Delta\baresq(s)}{\hatesq} \bigg]f_i^{\gamma\,(0)}
   + f_i^{\gamma\,(1)}(s)\;.
\end{equation}
Then, combining the above two equations,
\begin{subequations}
\label{u1-invariance}
\begin{eqnarray}
\nonumber
g_1^{\gamma}(s) & = & 1 - 
    \Bigg\{2\hatesq\frac{s}{\Lamsq}\Big(\fdw+\fdb\Big)\Bigg\}
  + \Bigg\{\hatgsq\frac{4\mwsq -s}{\Lamsq}\fdw 
          +\frac{3}{4}\hatgsq\frac{s}{\Lamsq}\fwww  \Bigg\}
\\ && \mbox{}\label{g1g-linear-a}
  - \Bigg\{ 4\hatgsq\frac{\mwsq}{\Lamsq}\fdw  \Bigg\}
  - \frac{s}{2\mwsq}\Bigg\{\frac{3}{2}\hatgsq\frac{\mwsq}{\Lamsq}\fwww
         - 6 \hatgsq \frac{\mwsq}{\Lamsq}\fdw  \Bigg\}
\\
\label{g1g-linear-b}
& = & 1 + 2 \hatgsq  \frac{s}{\Lambda^2} 
      \Big(\hatcsq f_{DW} - \hatssq f_{DB} \Big) \;.
\end{eqnarray}
\end{subequations}
The first term on the right-hand side of Eqn.~(\ref{g1g-linear-a}) is the 
tree-level value of $g_1$, 
$g_1^{\gamma\,(0)} = f_1^{\gamma\,(0)}$.  The second term is the contribution 
from the photon
self-energy, given by $\Delta\baresq(s)$, obtainable from 
Eqn.~(\ref{crctn_alpha}).  The third term is the direct correction to the 
three-point vertex, and the fourth term arises from the wave-function 
renormalization factor for the external $W$ bosons.  Notice that the constant 
pieces in these third and fourth terms cancel, as required by U(1)$_{\rm em}$
gauge invariance.  These first four terms comprise $f_1^{\gamma\,(\rm eff)}(s)$.
The last term of Eqn.~(\ref{g1g-linear-a}) is $f_2^{\gamma\,(\rm eff)}(s)$,
which receives only a direct correction.  Recall from Table~\ref{table-smff}
that $f_2^{\gamma\,(0)} = 0$.  In the final result, displayed in 
Eqn.~(\ref{g1g-linear-b}), the correction term is proportional to $s$, the 
square of the CM energy.  Indeed $g_1^\gamma(s)$ does reduce to its
canonical value, $g_1^\gamma(0)=1$, for on-shell photons.

However, some of the desirable properties of the SM \eeww amplitudes are 
not preserved in the amplitudes above.  In the SM elegant cancellations 
between the 
various Feynman graphs insure that the full amplitudes are well behaved at 
high energies, and perturbative unitarity is satisfied.  In particular, at 
high-energies the SM amplitudes behave like $s^n$ where $n\leq 0$, and, for 
large $\sqrt{s}$, the SM cross-sections decrease with increasing CM energy.  To 
the contrary, the amplitudes of this section will, in some cases, behave as 
$s^n$ where $n > 0$ leading to cross sections which do not decrease or even 
grow with increasing CM energy, violating tree-level perturbative 
unitarity at high energies.  As we approach the scale of the new interactions 
described by our effective Lagrangian, higher 
order terms in the expansion become increasingly important until the effective 
Lagrangian formalism breaks down. 

An explicit calculation, like the calculation leading to Eqn.~(\ref{amp-00}) and 
Eqn.~(\ref{amp-+0}), leads to the following high-energy limits for the 
amplitudes:
\begin{subequations}
\label{d6-high-energy}
\begin{eqnarray}
\label{d6-high-energy_+00}
{\cal M}^{\rm IB}(+;0,0) & \rightarrow &-\sqrt{2}\hatesq\gamma^2
   \frac{\mzsq}{\Lamsq} \bigg\{4\hat{s}^2\hatgzsq\fdb + \fb \bigg\} d^1_{+,0}
\;, \\ \label{d6-high-energy_-00} 
{\cal M}^{\rm IB}(-;0,0) & \rightarrow & \sqrt{2}\gamma^2\frac{\mzsq}{\Lamsq}
   \bigg\{2\hatcsq\hatgz^4\Big(\hat{c}^4\fdw + \hat{s}^4\fdb \Big)  
   + \frac{1}{2}\hatgsq\Big(\hatcsq\fw + \hatssq\fb\Big) \bigg\} d^1_{-,0}
\;, \\ 
\label{d6-high-energy_+pm0}
{\cal M}^{\rm IB}(+;\pm,0) & \rightarrow &-\sqrt{2}\hatesq\gamma
  \frac{\mzsq}{\Lamsq} \bigg\{4\hatssq\hatgzsq\fdb
  - \frac{1}{2}\Big(\fw-2\fbw-\fb\Big) \bigg\} 
   d^1_{+,\pm 1}
\;, \\ 
\label{d6-high-energy_+0mp}
{\cal M}^{\rm IB}(+;0,\mp) & = &{\cal M}^{\rm IB}(+;\pm,0)
\;, \\ 
\nonumber
{\cal M}^{\rm IB}(-;\pm,0) & \rightarrow &\sqrt{2}\gamma\frac{\mzsq}{\Lamsq}
   \bigg\{\hatcsq\hatgz^4\Big(-\hat{c}^4\fdw + 2\hat{s}^4\fdb \Big)
   + \frac{3}{4}\hatcsq\hatg^4\fwww 
\\ 
\label{d6-high-energy_-pm0} && \makebox[0cm]{}
	+ \frac{1}{4}\Big[\big(2\hatcsq - \hatssq\big)\fw
       +2\hatssq\fbw+\hatssq\fb\Big] \bigg\} 
   d^1_{-,\pm 1}
\;, \\ 
\label{d6-high-energy_-0mp}
{\cal M}^{\rm IB}(-;0,\mp) & = & {\cal M}^{\rm IB}(-;\pm,0) 
\;, \\ 
\label{d6-high-energy_-pmpm}
{\cal M}^{\rm IB}(-;\pm,\pm) & \rightarrow & \sqrt{2}\hatg^4\gamma^2
   \frac{\mwsq}{\Lamsq} \bigg\{ -6 \fdw + \frac{3}{2}\fwww\bigg\}
   d^1_{-,0} \;,
\end{eqnarray}
\end{subequations}
where $\gamma = E_W/m_W$.  The ${\cal M}^{\rm IB}(+;\pm,\pm)$ and 
${\cal M}^{\rm IB}(\tau;\pm,\mp)$ amplitudes do not receive any contributions 
that grow with energy.  Notice that $\fpone$ does not contribute to any of the 
above expressions.  We have used the equivalence theorem at the qualitative 
level to verify the behavior of these high-energy 
approximations\cite{equivalence}.

\subsection{The nonlinear realization of the symmetry-breaking sector}
\label{sec-eeww-nonlinear}

We now repeat the discussion for the chiral Lagrangian.
The corrections to the charge form factors, $\Delta \baresq(q^2)$, 
$\Delta \barssq(q^2)$, $\Delta \bargzsq(q^2)$ and $\Delta \bargwsq(q^2)$ may be 
found in Eqns.~(\ref{corrections_nl}). For the effective Lagrangian of 
Eqn.~(\ref{fulllagrangiannl}) the nonzero $f_i^{X\,(1)}(s)$ are
\begin{subequations}
\label{corr-dir-nl}
\begin{eqnarray}
\label{f3g-dir-nl}
f_3^{\gamma\,(1)}(s) & = & \hatgsq(-\alp{1}+\alp{2}+\alp{3}-\alp{8}+\alp{9})\;,
\end{eqnarray}
for the $WW\gamma$ vertex, and
\begin{eqnarray}
\label{f1z-dir-nl}
f_1^{Z\,(1)}(s) & = & \hatgzsq\alp{3} \;,\\
\label{f3z-dir-nl}
f_3^{Z\,(1)}(s) & = & f_3^{\gamma\,(1)}(s) 
	+ \hatgzsq(\alp{1} - \alp{2} + \alp{3}) \;,\\
\label{f5z-dir-nl}
f_5^{Z\,(1)}(s) & = & \hatgzsq\alp{11} \;,
\end{eqnarray}
\end{subequations}
for the $WWZ$ vertex. If $\Gamma_Z^{(0)}$ and $\Gamma_\gamma^{(0)}$ are the 
tree-level vertex functions and $\Gamma_Z^{(1)}$ and $\Gamma_\gamma^{(1)}$
are the one-loop vertex corrections, then we may define the `full' vertex 
functions according to $\hatgz\hatcsq\Gamma_Z^{(0)} \rightarrow 
\Delta \bargz(s) \hatcsq \Gamma_Z^{(0)} + \hatgz\Delta\barcsq(s) \Gamma_Z^{(0)}
+ \hatgz\hatcsq\Gamma_Z^{(1)} = \hatgz\hatcsq\Gamma_Z^{\rm (full)}$ and 
$\hate \Gamma_\gamma^{(0)} \rightarrow \Delta \bare(s) \Gamma_\gamma^{(0)}
+ \hate \Gamma_\gamma^{(1)} =  \hate \Gamma_\gamma^{\rm (full)}$; we find that 
$\Gamma_Z^{\rm (full)}$ and $\Gamma_\gamma^{\rm (full)}$ calculated in this 
way agree with the results of Ref.~\cite{aw93}.
Ref.~\cite{hol91} calculated the corrections to the $WW\gamma$
and $WWZ$ vertices for a small subset of the operators, but also discussed 
the \eeww cross-section.  

For the chiral Lagrangian the calculation of $g_1^\gamma$ is trivial.  There are 
no direct corrections from the $WW\gamma$ three-point vertex because 
$f_1^{Z\,(1)}(s) = f_2^{Z\,(1)}(s) = 0$.  There is no correction from the 
$W$-boson wave-function-renormalization factor because $Z_W = 1$.  Finally,
$\Delta \baresq(s) = 0$.  Therefore, $g_1^\gamma = 1$ respecting gauge 
invariance under U(1)$_{\rm em}$.

We present the high-energy limit of the \eeww amplitudes.
\begin{subequations}
\label{nl-high-energy}
\begin{eqnarray}
\label{nl-high-energy_+00}
{\cal M}^{\rm IB}(+;0,0) & \rightarrow & -2\sqrt{2}\hatesq\gamma^2
   \bigg\{\hatgzsq \alpha_2 \bigg\}  d^1_{+,0}
\;, \\ 
\label{nl-high-energy_-00}
{\cal M}^{\rm IB}(-;0,0) & \rightarrow & \sqrt{2}\hatgsq\gamma^2
   \bigg\{\hatgzsq \Big(\hatssq\alpha_2+\hatcsq\alpha_3\Big) + \hatgsq\alpha_9 
   \bigg\}  
   d^1_{-,0}
\;, \\ 
\label{nl-high-energy_+pm0}
{\cal M}^{\rm IB}(+;\pm,0) & \rightarrow & -\sqrt{2}\hatesq\gamma
  \bigg\{ \hatgzsq (\alpha_1+\alpha_2-\alpha_3\mp\alpha_{11})\bigg\}d^1_{+,\pm 1}
\;, \\ 
\label{nl-high-energy_+0mp}
{\cal M}^{\rm IB}(+;0,\mp) & = & {\cal M}^{\rm IB}(+;\pm,0)
\;, \\ 
\nonumber
{\cal M}^{\rm IB}(-;\pm,0) & \rightarrow & \sqrt{2}\hatgsq\gamma
   \bigg\{ \frac{1}{2}\hatgzsq \Big[ \hatssq(\alpha_1+\alpha_2-\alpha_3
   \mp\alpha_{11}) 
\\ 
\label{nl-high-energy_-pm0} &&\makebox[2cm]{}
+ \hatcsq(2\alpha_3 + \alpha_8+\alpha_9\pm\alpha_{11}) 
   \Big] \bigg\}d^1_{-,\pm 1}
\;, \\ 
\label{nl-high-energy_-0mp}
{\cal M}^{\rm IB}(-;0,\mp) & = & {\cal M}^{\rm IB}(-;\pm,0) 
\;.
\end{eqnarray}
\end{subequations}
The remaining amplitudes, ${\cal M}^{\rm IB}(\tau;\pm,\pm)$ and 
${\cal M}^{\rm IB}(\tau;\pm,\mp)$, do not have any contributions that grow with 
energy.  Like its counterpart $\fpone$ in the linear realization, $\beta_1$ 
makes no contribution which grows with energy.  Eqn.~(\ref{relate4}) may be used 
to verify that Eqns.~(\ref{d6-high-energy}) and Eqns.~(\ref{nl-high-energy})
are consistent under the equivalence of $\opone \sim \nl{1}^\prime$, 
$\obw \sim \nl{1}$, $\ob \sim \nl{2}$ and $\ow \sim \nl{3}$.  Notice that 
$\alpha_1$ does not contribute to the high-energy behavior of the 
$\lambda\overline{\lambda} = 00$ amplitudes as was 
observed in Ref.~\cite{hol91}.  On the other hand, we observe that $\alpha_1$
makes a contribution to the ${\cal M}^{\rm IB}(\tau;\pm,0)$ and 
${\cal M}^{\rm IB}(\tau;0,\pm)$ amplitudes which is proportional to $\gamma$.

\subsection{The `phenomenological Lagrangian'}
\label{sec-eeww-nabgbv}

To facilitate discussion and to make a connection with much of the standard 
literature, we also present a phenomenological Lagrangian of the most 
general $WW\gamma$ and $WWZ$ couplings which respects only U(1)$_{\rm em}$ 
gauge invariance\cite{hpzh87}.  Retaining only those terms that are CP 
conserving,
\begin{eqnarray}
\nonumber
\lefteqn{ \makebox[-1.5cm]{}{\cal L}_{WWV} =  - i g_{WWV} \Bigg\{ 
g_1^V \Big( W^+_{\mu\nu} W^{- \, \mu} V^{\nu} 
  - W^+_{\mu} V_{\nu} W^{- \, \mu\nu} \Big) 
+ \kappa_V W_\mu^+ W_\nu^- V^{\mu\nu}}
\\ && \mbox{} 
+ \frac{\lambda_V}{m_W^2} W^+_{\mu\nu} W^{- \, \nu\rho} V_\rho^{\; \mu}
- i g_5^V \epsilon^{\mu\nu\rho\sigma} \Big[ W^+_\mu ( \partial_\rho W^-_\nu ) -
( \partial_\rho W^+_\mu ) W^-_\nu \Big] V_\sigma \Bigg\}
\;,\label{lagr_phenom}
\end{eqnarray}
where the overall coupling constants are $\hat{g}_{WW\gamma} = \hate$ and 
$\hat{g}_{WWZ} = \hatgz \hatcsq$.  The field-strength tensors include only
the Abelian parts, \ie\/ $W^{\mu\nu} = \partial^\mu W^\nu - \partial^\nu W^\mu$
and $V^{\mu\nu} = \partial^\mu V^\nu - \partial^\nu V^\mu$.  
The explicit relationships between the form factors $f_i^V$ 
and the effective Lagrangian of Eqn.~(\ref{lagr_phenom}) are given by
\begin{subequations}
\label{rosetta}
\begin{eqnarray}
f_1^V(s) & = & g_1^V + \frac{s}{2 m_W^2} \lambda_V \;,\\
f_2^V(s) & = & \lambda_V \;,\\
f_3^V(s) & = & g_1^V + \kappa_V + \lambda_V \;,\\
f_5^V(s) & = & g_5^V \;.
\end{eqnarray}
\end{subequations}

The \eeww amplitudes with corrections from Eqn.~(\ref{lagr_phenom}) display 
the following high-energy behavior.
\begin{subequations}
\label{nabgbv-high-energy}
\begin{eqnarray}
\label{nabgbv-high-energy_+00}
{\cal M}^{\rm IB}(+;0,0) & \rightarrow & -2\sqrt{2}\hatesq\gamma^2
  \bigg\{\Delta\kappa_\gamma-\Delta\kappa_Z\bigg\} d^1_{+,0}
\;, \\ 
\label{nabgbv-high-energy_-00}
{\cal M}^{\rm IB}(-;0,0) & \rightarrow & \sqrt{2}\hatgsq\gamma^2
  \bigg\{2\hatssq\Delta\kappa_\gamma + \Big(\hatcsq-\hatssq \Big)\Delta\kappa_Z 
  \bigg\} d^1_{-,0}
\;, \\ 
\nonumber
{\cal M}^{\rm IB}(+;\pm,0) & \rightarrow & -\sqrt{2}\hatesq\gamma
\bigg\{\Big(\Delta\kappa_\gamma +\lambda_\gamma + \Delta g_1^\gamma \pm 
       g_5^\gamma \Big)
\\ \label{nabgbv-high-energy_++0} && \makebox[1.5cm]{}
       - \Big( \Delta\kappa_Z + \lambda_Z + \Delta g_1^Z \pm g_5^Z \Big) 
      \bigg\} d^1_{+,\pm 1}
\;, \\ 
\label{nabgbv-high-energy_+0mp}
{\cal M}^{\rm IB}(+;0,\mp) & = & {\cal M}^{\rm IB}(+;\pm,0)
\;, \\ 
\nonumber
{\cal M}^{\rm IB}(-;\pm,0) & \rightarrow & \sqrt{2}\hatgsq\gamma
  \bigg\{\hatssq\Big(\Delta\kappa_\gamma+\lambda_\gamma+\Delta g_1^\gamma 
  \pm g_5^\gamma\Big) 
\\ \label{nabgbv-high-energy_-+0} & & \makebox[2cm]{}+\frac{1}{2}
  \Big(\hatcsq-\hatssq \Big)\Big(\Delta\kappa_Z +\lambda_Z+\Delta g_1^Z 
  \pm g_5^Z \Big)
  \bigg\} d^1_{-,\pm 1}
\;, \\ 
\label{nabgbv-high-energy_-0mp}
{\cal M}^{\rm IB}(-;0,\mp) & = & {\cal M}^{\rm IB}(-;\pm,0) 
\;, \\ 
\label{nabgbv-high-energy_+pmpm}
{\cal M}^{\rm IB}(+;\pm,\pm) & = & -2\sqrt{2}\hatesq\gamma^2
  \bigg\{\lambda_\gamma-\lambda_Z \bigg\} d^1_{+,0}
\;, \\ 
\label{nabgbv-high-energy_-pmpm}
{\cal M}^{\rm IB}(-;\pm,\pm) & = &  \sqrt{2}\hatgsq\gamma^2
  \bigg\{2\hatssq\lambda_\gamma+ \Big(\hatcsq-\hatssq \Big)\lambda_Z \bigg\} 
  d^1_{-,0}
\;.
\end{eqnarray}
\end{subequations}
The remaining amplitudes, ${\cal M}^{\rm IB}(\tau;\pm,\mp)$, do not have any 
contributions that grow with energy.  Here $\Delta\kappa_V = \kappa_V - 1$
and $\Delta g_1^V = g_1^V - 1$ for $V = \gamma,Z$.


\section{Numerical Analysis}\label{sec-numerical}

In this section we shall determine the level to which the parameters of the 
effective Lagrangians (\ref{fulllagrangian}), (\ref{fulllagrangiannl}) and 
(\ref{lagr_phenom}) may be measured/constrained through the study of $W$-boson
pair production at LEP II and at a 500GeV linear collider.  We are especially 
interested in comparing and contrasting the results which we obtain in the 
different realizations of the symmetry-breaking sector. 

When analysing actual experimental data $W$-boson finite-width 
effects\cite{aow94,lepii-ww-wg,width,lepii-tgc-wg} and contributions from 
initial and final state radiation\cite{lepii-tgc-wg,ifrad} are very important.
However, as verified by Ref.~\cite{lepii-tgc-wg}, these contributions 
primarily lead to a shift in the measured quantities, but the sensitivity to 
non-standard couplings is minimally affected.  Hence, we may justifiably 
use the simplified calculation of Section~\ref{sec-corrections}.

The calculation of the cross section for \eeww\/, $W^- \rightarrow f_1 
\overline{f}_2$, $W^+ \rightarrow f_3 \overline{f}_4$ requires, in general,
the evaluation of an eight-dimensional integral.  In 
Sec.~\ref{sec-formfactor} we introduced $\Theta$, the angle between the 
momentum vectors of the $W^-$ and the $e^-$ as measured in the CM frame.
The integration over the azimuthal angle of the $W$-boson momentum vectors, 
$\Phi$, is trivial, and need not be considered explicitly.  We do not consider 
transverse polarizations of the LEP~II beams\cite{zep87}. In the zero-width 
approximation for the decaying $W$ bosons, two integrations are performed 
analytically, and a single event is characterized by five angles.  Using the 
same notation as Ref.~\cite{hpzh87} we introduce the momentum vectors of $f_1$ 
and $\overline{f}_2$ as measured in the rest frame of the $W^-$ as 
\begin{subequations}
\label{Wminus-decay}
\begin{eqnarray}
\label{f1-momentum}
p_1^\mu & = & \frac{1}{2}\sqrt{s}\Big( 1, \sin\theta\cos\phi, 
                      \sin\theta\sin\phi, 
                      \cos\theta\Big)\;,\\
\label{f2-momentum}
p_2^\mu & = & \frac{1}{2}\sqrt{s}\Big( 1, -\sin\theta\cos\phi, 
                      -\sin\theta\sin\phi, 
                      -\cos\theta\Big)\;.
\end{eqnarray}
\end{subequations}
For the momentum vectors of $f_3$ and $\overline{f}_4$ as measured in the rest 
frame of the $W^+$ we have 
\begin{subequations}
\label{Wplus-decay}
\begin{eqnarray}
\label{f3-momentum}
p_3^\mu & = & \frac{1}{2}\sqrt{s}\Big( 1, 
                      -\sin\overline{\theta}\cos\overline{\phi}, 
                      -\sin\overline{\theta}\sin\overline{\phi}, 
                      -\cos\overline{\theta}\Big)\;,\\
\label{f4-momentum}
p_4^\mu & = & \frac{1}{2}\sqrt{s}\Big( 1, 
                      \sin\overline{\theta}\cos\overline{\phi}, 
                      \sin\overline{\theta}\sin\overline{\phi}, 
                      \cos\overline{\theta}\Big)\;.
\end{eqnarray}
\end{subequations}
The z-axis and the x--z plane are common to all three frames.

In practice, to perform a fit, we need to retain as much of the above angular 
information as possible.  A straightforward approach to this problem is to 
compare the full five-fold differential cross-section calculated in the SM
to that calculated with the nonstandard contributions.  Suppose, for a moment, 
that individual events could be completely reconstructed.  Suppose also that we 
divide each of the above angular variables into 10 bins.  Then in total we have 
$10^5$ bins to consider; at LEP~II, where approximately $8\times 10^3$ $W$-pair 
events are expected, we can expect to have zero events in a very large number of 
these bins.  Hence, we will use many fewer bins.  In particular we employ 
four bins in each variable for $4^5 = 1,024$ total bins, and we then perform a 
log-likelihood fit with Poisson statistics\cite{pdg96}.  A similar strategy 
was employed by the authors of Ref.~\cite{cgg95andgg96}.

The most common final state, realized in 49\% of the events\cite{lepii-tgc-wg}, 
is where both $W$ bosons decay hadronically, \ie\/ the $jjjj$ final state.  
For many of these events 
it is possible to reconstruct the four-momenta of all four jets.  The jets may
then be paired such that each pair has the invariant mass of an on-shell 
$W$-boson.  However, it is extremely difficult within each pair to determine
which jet came from a quark and which came from an anti-quark.  
By tagging charm quarks it may be possible to make
correct assignments of the jets in some fraction of the events.  Color 
reconnection effects\cite{lepii-mw-wg,color} may be 
important at LEP~II where the $W$ bosons are produced with a very small 
velocity, and hence their production and decay vertices are minimally 
displaced, and as a consequence their decay jets may interact between pairs.  
Following Ref.~\cite{lepii-tgc-wg} we assume that 
there is an overall ambiguity in the assignment of the jets within each 
pair, and we also assume that we cannot determine the charges of the $W$ 
bosons.  In this respect our analysis is somewhat conservative.  Again 
following Ref.~\cite{lepii-tgc-wg}, we assign an efficiency of 60\% for the 
reconstruction of the $jjjj$ final state.

Next we consider the final state where one $W$ boson decays hadronically, and the
the other decays leptonically, \ie\/ the $jjl\nu$ final state.  The branching
fraction is 14\% for $l=e$, 14\% for $l=\mu$ and 14\% for $l=\tau$.  Due to 
difficulties in the reconstruction we will simply ignore the $jj\tau\nu$ final
state.  For the reconstruction of the $jjl\nu$, $l=e,\mu$ final states, only 
one reconstruction ambiguity exists; it is difficult to correctly determine 
which of the jets is the quark jet and which is the anti-quark jet.  Because
the charge assignments of the $W$ bosons are determined by the measurement 
of the lepton charge, charm-quark tagging might be useful for assigning the 
jets in some portion of the events; for simplicity will we ignore this 
refinement.  Following Ref.~\cite{lepii-tgc-wg}, we assign an efficiency of 95\% 
for the reconstruction of the $jjl\nu$ final state.

Finally, there is the final state where both $W$ bosons decay leptonically, 
\ie\/ the $l\nu l^\prime\nu^\prime$ final state, which occurs 9\% of the time.  
A portion of these events, where one or both of the leptons is a $\tau$ lepton, 
is difficult to reconstruct.  The remaining events may be well reconstructed up 
to an overall two-fold ambiguity which is the result of having two neutrinos 
in the final state.  We have chosen to neglect the $l\nu l^\prime\nu^\prime$ 
final state in this analysis, but it may be straightforwardly added to future 
analyses.

We make one kinematical cut, $|\cos\Theta|<0.9$, and divide each of the five 
variables into four bins.  In a more complete analysis we would need to add 
additional separation cuts on the final-state fermions, especially to allow 
for a jet-cone radius.  These cuts are crudely included through the
incorporation of the efficiencies.

For the one-sigma limits on the coefficients of the 
energy-dimension-six operators in the linear realization of the 
symmetry-breaking sector see Table~\ref{table-d6_limits}.
\begin{table}[htb]
\begin{tabular}{|c||c|c|c|c|c|c|c|}
& $f_{DW}$ & $f_{DB}$ & $f_{BW}$ & $f_{\Phi,1}$ & $f_{WWW}$ & $f_{W}$ & $f_{B}$ 
\\ \hline\hline
LEP~II & 2.1   & 12   & 1.5  & 0.19  & 10   & 7.1  & 46 
\\ \hline 
LC     & 0.063 & 0.39 & 0.32 & 0.045 & 0.23 & 0.10 & 0.25
\\ 
\end{tabular}
\caption{One-sigma limits on the parameters of the linearly realized effective
Lagrangian assuming $\Lambda = 1$TeV.  In the first row are the 
constraints from LEP~II at 175GeV with ${\cal L}^{\rm int} = 500{\rm pb}^{-1}$, 
and the second row contains results for a 500GeV future linear collider with 
${\cal L}^{\rm int} = 50{\rm fb}^{-1}$.  The one-sigma allowed region is 
approximately symmetric about zero.}
\label{table-d6_limits}
\end{table}
We include results not only for LEP~II with  ${\cal L}^{\rm int}
= 500{\rm pb}^{-1}$ at $\sqrt{s} = 175{\rm GeV}$, but we also perform the 
analysis for a 
future linear collider with $\sqrt{s} = 500{\rm GeV}$ and ${\cal L}^{\rm int}
= 50{\rm fb}^{-1}$.  We repeat the analysis for the parameters of the effective 
Lagrangian in the scenario where symmetry breaking is realized nonlinearly.  The 
results are presented in Table~\ref{table-nl_limits}.
\begin{table}[htb]
\begin{tabular}{|c||c|c|c|c|c|c|c|}
& $\beta_1$ & $\alpha_1$ &  $\alpha_2$ &  $\alpha_3$ &  
              $\alpha_8$ &  $\alpha_9$ &  $\alpha_{11}$ 
\\ \hline\hline
LEP~II & 0.0028  & 0.022  & 0.34   & 0.053   & 0.017  & 0.10    & 0.50
\\ \hline 
LC     & 0.00064 & 0.0047 & 0.0018 & 0.00072 & 0.0022 & 0.00078 & 0.0045 
\\ 
\end{tabular}
\caption{One-sigma limits on the parameters of the nonlinearly realized effective
Lagrangian.  In the first row are the constraints from LEP~II at 175GeV with 
${\cal L}^{\rm int} = 500{\rm pb}^{-1}$, 
and the second row contains results for a 500GeV future linear collider with 
${\cal L}^{\rm int} = 50{\rm fb}^{-1}$.  The one-sigma region is approximately 
symmetric about zero.}
\label{table-nl_limits}
\end{table}
The results for the analysis in the basis of Eqn.~(\ref{lagr_phenom}) appear in 
Table~\ref{table-nabgbv_limits}.
\begin{table}[htb]
\begin{tabular}{|c||c|c|c|c|c|c|c|c|}
& $\Delta g_1^\gamma$ & $\Delta g_1^Z$ & 
  $\Delta \kappa_\gamma$ & $\Delta \kappa_Z$ &
  $\lambda_\gamma$ & $\lambda_Z$ & $f_5^\gamma$ &  $f_5^Z$
\\ \hline\hline
LEP~II & 0.12   & 0.073  & 0.092   & 0.067   & 0.11   & 0.068  & 0.48 & 0.28
\\ \hline 
LC     & 0.0030 & 0.0030 & 0.00059 & 0.00077 & 0.0022 & 0.0017 & 
$\mbox{}^{+0.0036}_{-0.0023}$ & 0.0025
\\ 
\end{tabular}
\caption{One-sigma limits on the couplings from the phenomenological effective 
Lagrangian of Eqn.~(\protect\ref{lagr_phenom}).  In the first row are the 
constraints from LEP~II at 175GeV 
with ${\cal L}^{\rm int} = 500{\rm pb}^{-1}$, 
and the second row contains results for a 500GeV future linear collider with 
${\cal L}^{\rm int} = 50{\rm fb}^{-1}$.  The one-sigma region is approximately 
symmetric about zero  except for $f_5^\gamma$.}
\label{table-nabgbv_limits}
\end{table}
We report the one-sigma limits to two significant digits, even though the 
second digit is only approximate.  In many 
cases we find that one-sigma region is not perfectly symmetric about zero, 
but generally the asymmetry is less than 10\%.

We performed several cross checks of our results.  First of all, for the 
LEP~II constraints on $f_{WWW}$, $f_W$ and $f_B$, see 
Table~\ref{table-d6_limits}, we were able to make some comparisons with the 
results of Ref.~\cite{lepii-tgc-wg}; we found good agreement.  For a few of 
the parameters in Table~\ref{table-nl_limits} and Table~\ref{table-nabgbv_limits}
we were able to compare with the results of Refs.~\cite{cgg95andgg96}, again 
finding good agreement.  Additionally we made several checks for the internal 
consistency of our results, for example, by using the relationships of 
Eqns.~(\ref{relate4}).  Additional relationships connect some of the values in 
Table~\ref{table-d6_limits} and Table~\ref{table-nl_limits} to those in 
Table~\ref{table-nabgbv_limits}.  Unfortunately the high-energy
limits presented in Eqns.~(\ref{d6-high-energy}) for the light-Higgs scenario,
in Eqns.~(\ref{nl-high-energy}) for the chiral Lagrangian and in 
Eqns.~(\ref{nabgbv-high-energy}) for the phenomenological effective Lagrangian
are not useful for explaining the improvement from LEP~II to the linear 
collider.  This is simply because LEP~II is much too close to the $W$-boson
pair production threshold for the high-energy approximation to be useful.


\section{Discussion}\label{sec-discussion}

Compare the first row of Table~\ref{table-d6_limits}, obtained from 
studying $e^+e^- \rightarrow W^+W^-$, to the constraints in Eqn.~(\ref{fit_now}),
obtained from studying the low-energy and $Z$-pole data.  The first observation 
we make is that, current constraints on $f_{DW}$ are sufficiently strong that
we do not have sensitivity to this parameter at LEP~II.  On the other hand,
the bounds on $f_{DB}$ and $f_{\Phi,1}$ here are only slightly
weaker than current bounds.  Hence, with the improvements to the analysis 
described in the previous section, perhaps the improved bounds on these two 
coefficients may become competitive.  Finally, the new bound on $f_{BW}$
is an improvement over the current bound.  The current data is 
not sufficient to rule out observable effects from the operator ${\cal O}_{BW}$, 
contrary to some expectations\cite{drghm92}.  However, next consider the 
constraints of Eqn.~(\ref{fit_lep2}), obtained from studying $e^+e^- \rightarrow 
f\overline{f}$ at LEP~II.  Immediately we see that, if any of the four 
coefficients 
$f_{DW}$, $f_{DB}$, $f_{BW}$ or $f_{\Phi,1}$ were to make an observable 
contribution at LEP~II to $e^+e^- \rightarrow W^+W^-$, then there would be 
an even larger effect observed in the $e^+e^- \rightarrow f\overline{f}$ 
channel.  Hence, if measurements made on two-fermion final states are in good
agreement with the SM, then we may disregard these four coefficients, and we
are justified in considering only $f_{WWW}$, $f_{W}$ and $f_{B}$ when studying 
$W$-boson pair production.  In this case we may employ the 
relations\cite{hisz93}
\begin{subequations}
\label{hisz-relations}
\begin{eqnarray}
\label{g1z-hisz}
g^Z_1(q^2) & = & 1 + \frac{1}{2}\frac{m_Z^2}{\Lambda^2}f_W \;, \\
\label{kappagamma-hisz}
\kappa_\gamma (q^2) & = & 1 + \frac{1}{2}\frac{m_W^2}{\Lambda^2}
  \Big(f_W + f_B\Big) \;, \\
\label{kappaz-hisz}
\kappa_Z(q^2) & = & 1 + \frac{1}{2}\frac{m_Z^2}{\Lambda^2}
  \Big(\hatcsq f_W - \hatssq f_B\Big)\;, \\
\label{lambda-hisz}
\lambda_{\gamma}(q^2)  =  \lambda_Z(q^2)
& = & 
  \frac{3}{2}\hatgsq \frac{m_W^2}{\Lambda^2} f_{WWW} \;.
\end{eqnarray}
\end{subequations}
Out of the three parameters $g^Z_1$, $\kappa_\gamma$ and $\kappa_Z$, only two 
are independent.  Notice that $g^\gamma_1 = 1$.  Similarly, only one of the 
$\lambda$ couplings is independent.

Next, consider the second row of Table~\ref{table-d6_limits}.  We see that
the bounds on all seven of the parameters improve at a 500GeV linear collider.
In some cases, such as the constraints of $f_{\Phi,1}$, we expect small 
improvements due to the high luminosity which more than compensates for the 
$1/s$ falloff in the cross-section.  Approximately we might expect an 
improvement from statistics roughly of the order 
$\sqrt{N_{\rm LC}}/\sqrt{N_{\rm LEP~II}} 
\approx \sqrt{{\cal L}_{\rm LC}/s_{\rm LC}}/
\sqrt{{\cal L}_{\rm LEP~II}/s_{\rm LEP~II}} \approx 3.5$; we see an improvement 
in the measurement of $f_{\Phi,1}$ by approximately a factor of $4$.  However, 
from studying 
Eqns.~(\ref{d6-high-energy}), we see that the $f_i$'s often appear multiplied 
by factors of $\gamma$ or $\gamma^2$, which, upon modification of the above 
argument, suggest an improvement due to statistics by a factor of 6 or 10
respectively.  We see that these estimates tend to fail because the LEP~II CM 
energy is too low for the high-energy approximations of the amplitudes to be 
useful.

Next we compare the linear-collider constraints on $f_{DW}$, $f_{DB}$, 
$f_{BW}$ and $f_{\Phi,1}$ from Table~\ref{table-d6_limits} with 
Eqn.~(\ref{fit_nlc}), and we see that, if there is a signal from these 
coefficients 
in the $e^+e^- \rightarrow W^+W^-$ process, then there should be an even bigger
signal in the $e^+e^- \rightarrow f\overline{f}$ channel.  Hence, if we see 
no signal in the latter channel, then we can return to the three parameter 
fit in terms of $f_{WWW}$, $f_{W}$ and $f_{B}$, and Eqns.~(\ref{hisz-relations})
may be employed.

Now we turn to the chiral Lagrangian.  Recall that $\beta_1$, $\alpha_1$ and
$\alpha_8$ are already constrained by the low-energy data through their 
contributions to the gauge-boson two-point-functions.  One of these parameters,
$\alpha_1$, also contributes directly to the $WW\gamma$ and $WWZ$ vertices.
If we could justify neglecting these three parameters, then we can present a set
of relations that parallels Eqns.~(\ref{hisz-relations}):
\begin{subequations}
\label{nl-hisz-relations}
\begin{eqnarray}
\label{nl-g1z-hisz}
g_1^Z(q^2) & = & 1 + \hatgzsq \alpha_3 \;,\\
\label{nl-kappagamma-hisz}
\kappa_\gamma(q^2) & = & 1 + 
\hatgsq \Big( \alpha_2 + \alpha_3 + \alpha_9 \Big)\;, \\
\label{nl-kappaz-hisz}
\kappa_Z(q^2) & = & 1 + \hatgzsq \Big( -\hatssq\alpha_2  + \hatcsq\alpha_3 
+ \hatcsq\alpha_9 \Big)  \;, \\
\label{nl-g5z-hisz}
g_5^Z(q^2) & = & \hatgzsq \alpha_{11} \;,\\
\label{nl-lambda-hisz}
\lambda_{\gamma}(q^2)  \approx  \lambda_Z(q^2)
& \approx & 0 \;.
\end{eqnarray}
\end{subequations}
These results agree with Ref.~\cite{aw93}.  The numerical estimate of 
Eqn.~(\ref{aw-est-alpha9}) suggests that $\alpha_9$ may be very small.  If we 
neglect $\alpha_9$, then, upon using Eqns.~(\ref{relate4}),
Eqns.~(\ref{nl-g1z-hisz})-(\ref{nl-kappaz-hisz}) are equivalent to 
Eqns.~(\ref{g1z-hisz})-(\ref{kappaz-hisz}); in general we must retain 
$\alpha_9$.  The appearance of $g_5^Z$ at the 
leading order in the chiral Lagrangian has no leading-order counterpart in the 
light-Higgs 
scenario.  By Eqn.~(\ref{aw-est-alpha11}) we expect that $\alpha_{11}$, hence
$g_5^Z$, may be
small, again due to custodial symmetry.  However, its suppression is not so 
strong as for the other custodial-symmetry-violating couplings.  Also, because
it has no counterpart in the light-Higgs scenario, it is of special interest for 
discriminating between the two realizations of the symmetry-breaking sector.
Finally, because the $\lambda$ couplings are inherently higher order in the 
chiral Lagrangian, Eqn.~(\ref{nl-lambda-hisz}) is obtained trivially at low
energies.

Upon comparing the first row of Table~\ref{table-nl_limits} with the 
$Z$-pole/low-energy constraints of Eqn.~(\ref{fit_now_nl}), we see that, 
at LEP~II, we are justified in neglecting the contributions of $\beta_1$,
$\alpha_1$ and $\alpha_8$, hence Eqns.~(\ref{nl-hisz-relations}) are valid.  
Also, if we believe the estimates of Eqns.~(\ref{chiral-quark-est}), then we 
need to constrain the $\alpha$ parameters at the level of $10^{-3}$ before 
we can expect to see the effects of new physics.  Clearly we do not yet have 
this type of sensitivity at LEP~II.

Considering the second row of Table~\ref{table-nl_limits}, we expect that the
linear collider may be sensitive to new physics described by the chiral 
Lagrangian.  However, the analysis now becomes more complicated.  
Linear-collider experiments may also be sensitive to $\alpha_8$ and marginally
sensitive to $\alpha_1$.  In stark contrast to the light-Higgs scenario, with 
the chiral Lagrangian we do not obtain additional constraints by studying
$e^+e^- \rightarrow f\overline{f}$.  Here the leading corrections to the 
gauge-boson propagators are independent of $q^2$, and hence there is no 
benefit from the higher CM energy.  To the contrary, once we are away from the 
$Z$ pole, event rates are low and we are statistics limited.  If, 
taking advantage of the high luminosity of the linear collider, we repeat the 
LEP experiments on the $Z$ pole, then, through the improved measurement of 
$\barssq(\mzsq)$, it may be possible to improve the measurements of $\beta_1$, 
$\alpha_1$ and $\alpha_8$ directly.  The measurement  of the weak mixing angle
may also be improved at the TeV33.  However, the impact of these additional
measurements is limited.

Finally, in Table~\ref{table-nabgbv_limits}, we have presented constraints 
which treat corrections to three-gauge-boson vertices as independent from 
the two-point-function corrections.  As we see from Eqn.~(\ref{nl-g5z-hisz}),
we are justified in measuring $f_5^Z$ (which is equivalent to $g_5^Z$) 
separately from the rest.  Eqn.~(\ref{lambda-hisz}) implies that 
$\lambda_\gamma = \lambda_Z < 0.04$ at LEP~II, and $\lambda_\gamma = \lambda_Z 
< 0.001$ at the linear collider; these results should be contrasted with the 
first-row and second-row results of Table~\ref{table-nabgbv_limits} 
respectively.  The other correlations described by Eqns.~(\ref{hisz-relations})
and Eqns.~(\ref{nl-hisz-relations}) could also be explored in this way.  
However, at linear collider energies where the sensitivity to $WW\gamma$ and 
$WWZ$ couplings rivals the sensitivity to gauge-boson propagator corrections, 
it is more sensible to abandon the analysis of Table~\ref{table-nabgbv_limits} 
in favor of the analyses of Table~\ref{table-d6_limits} and 
Table~\ref{table-nl_limits}.


\section{Conclusions}\label{sec-conclusions}

When we consider the effects of new physics described by an effective 
Lagrangian with the linearly realized symmetry-breaking sector, {\em i.e.} 
the light Higgs scenario, then, at the leading order, the coefficients of 
seven operators contribute to $e^+e^- \rightarrow W^+W^-$ amplitudes.  These 
coefficients are $f_{DW}$, $f_{DB}$, $f_{BW}$, $f_{\Phi,1}$, $f_{WWW}$, 
$f_{W}$ and $f_{B}$.  The first four are already constrained via low-energy 
and $Z$-pole experiments, but the current constraints, in some cases, do not 
rule out observable contributions to $e^+e^- \rightarrow W^+W^-$ at LEP~II.  
However, the constraints on these four coefficients may be strengthened by 
also studying $e^+e^- \rightarrow f\overline{f}$ processes at LEP~II.  In 
fact, the $e^+e^- \rightarrow f\overline{f}$ process is more sensitive to 
these four parameters than is the $e^+e^- \rightarrow W^+W^-$ process.  Hence, 
if we fail to observe a signal for non-SM physics in the $e^+e^- \rightarrow 
f\overline{f}$ channel, then we can neglect these four coefficients when 
analysing \eeww.\/  The analysis of $e^+e^- \rightarrow W^+W^-$ amplitudes 
then reduces to a three-parameter analysis in terms of $f_{WWW}$, $f_{W}$ and 
$f_{B}$.  The analysis may be performed using the familiar parameters 
$\kappa_V$, $g_1^V$ and $\lambda_V$ with $V = \gamma$, $Z$ subject to the 
constraints of Eqn.~(\ref{hisz-relations}).
Essentially the same scenario occurs at the linear collider.  We must use 
both the $W$-boson pair-production process as well as studies of two-fermion
final states to separate the contributions of $f_{DW}$, $f_{DB}$, $f_{BW}$ and 
$f_{\Phi,1}$ from those of $f_{WWW}$, $f_{W}$ and $f_{B}$.

When we consider the effects of new physics described by an effective 
Lagrangian with the symmetry-breaking realized nonlinearly, {\em i.e.} the 
chiral Lagrangian which does not include a physical Higgs scalar boson, we 
must consider the contributions of $\beta_{1}$, $\alpha_{1}$, $\alpha_{2}$, 
$\alpha_{3}$, $\alpha_{8}$, $\alpha_{9}$ and $\alpha_{11}$.  Three of these, 
$\beta_1$, $\alpha_{1}$ and $\alpha_{8}$, are already constrained by the 
low-energy and $Z$-pole data.  Hence Eqns.~(\ref{nl-hisz-relations}) may be 
relevant.  The inclusion of $\alpha_9$ makes these relations slightly 
more complicated than their light-Higgs-scenario counterparts, 
Eqns.~(\ref{hisz-relations}).  The parity violating coupling $\alpha_{11}$ 
also contributes, but certainly we can disentangle its effects by 
constructing some parity-violating observables.  Because it has no 
leading-order parity-violating counterpart in the light-Higgs scenario, 
$\alpha_{11}$ is especially interesting.  
At the linear collider $e^+e^- \rightarrow W^+W^-$ amplitudes may 
be sensitive to all seven parameters, providing for a rather complicated 
analysis.  

We also presented an analysis where the most general contributions to the 
$WW\gamma$ and $WWZ$ vertices are assumed to be independent of the corrections
to the gauge-boson two-point-functions.  At LEP~II this analysis is useful,
especially for testing the relations of Eqns.~(\ref{hisz-relations}).
However, at a 500GeV linear collider, where the measurements of $WW\gamma$ 
and $WWZ$ couplings become competitive with measurements of gauge-boson
propagator corrections, this approach may be less useful.


\section*{Acknowledgements}
The authors would like to thank Dieter Zeppenfeld, Seong-Youl Choi, Bernd Kniehl,
Ben Bullock, Masaharu Tanabashi and Mihoko Nojiri for helpful discussions.  We 
would also like to thank Seiji~Matsumoto for providing us with the results of 
the recent global analysis of precision data.


\end{document}